\PassOptionsToPackage{table}{xcolor}
\documentclass[a4paper, amsfonts, amssymb, amsmath, reprint, showkeys, nofootinbib, twoside, superscriptaddress]{revtex4-1}
\usepackage[english]{babel}
\usepackage[utf8]{inputenc}
\usepackage{comment}
\usepackage{array}
\usepackage[acronym]{glossaries}
\usepackage{tikz}
\usetikzlibrary{quantikz2}
\usepackage{enumitem}

\usepackage{physics}
\usepackage{bm}
\usepackage{mathtools}
\usepackage{placeins}
\usepackage{enumitem}
\usepackage{subfigure}

\usepackage[colorlinks=true, urlcolor=blue, linkcolor=blue, citecolor=blue]{hyperref}

\usepackage{multirow}
\newacronym{PES}{PES}{potential energy surface}
\newacronym{VSCF}{VSCF}{vibrational self-consistent field}
\newacronym{QFT}{QFT}{quantum Fourier transform}
\newacronym{CSA}{CSA}{Cartan subalgebra}
\newacronym{HO}{HO}{harmonic oscillator}
\newacronym{CGF}{CGF}{Christiansen Greedy Fragments}
\newacronym{FC}{FC}{Fully commuting}
\newacronym{QWC}{QWC}{Qubit-wise commuting}
\newacronym{PF}{PF}{Pauli fragments}
\newacronym{BF}{BF}{Bosonic fragments}
\newacronym{RS}{RS}{real space}
\newacronym{FT}{FT}{Fourier transform}
\newacronym{ZPE}{ZPE}{zero point energy}
\newacronym[plural={occupation number vectors}, shortplural={ONVs}]{ONV}{ONV}{occupation number vector}
\newacronym{QPE}{QPE}{quantum phase estimation}

\DeclareMathOperator*{\argmax}{argmax}
\newcommand{\bea}{\begin{eqnarray}}
\newcommand{\eea}{\end{eqnarray}}
\newcommand{\eq}[1]{Eq.~(\ref{#1})} %
\begin{document}
\title{Trotter simulation of vibrational Hamiltonians on a quantum computer}

\author{Shreyas Malpathak}
\affiliation{Department of Physical and Environmental Sciences,
University of Toronto Scarborough, Toronto, Ontario M1C 1A4, Canada}
\affiliation{Chemical Physics Theory Group, Department of Chemistry,
University of Toronto, Toronto, Ontario M5S 3H6, Canada}
\author{Sangeeth Das Kallullathil}
\affiliation{Department of Physical and Environmental Sciences,
University of Toronto Scarborough, Toronto, Ontario M1C 1A4, Canada}
\affiliation{Chemical Physics Theory Group, Department of Chemistry,
University of Toronto, Toronto, Ontario M5S 3H6, Canada}
\author{Ignacio Loaiza}
\affiliation{Xanadu, Toronto, ON, M5G 2C8, Canada}
\author{Stepan Fomichev}
\affiliation{Xanadu, Toronto, ON, M5G 2C8, Canada}
\author{Juan Miguel Arrazola}
\affiliation{Xanadu, Toronto, ON, M5G 2C8, Canada}
\author{Artur F. Izmaylov}
\affiliation{Department of Physical and Environmental Sciences,
University of Toronto Scarborough, Toronto, Ontario M1C 1A4, Canada}
\affiliation{Chemical Physics Theory Group, Department of Chemistry,
University of Toronto, Toronto, Ontario M5S 3H6, Canada}

\date{\today}

\begin{abstract}

Simulating vibrational dynamics is essential for understanding molecular structure, unlocking useful applications such as vibrational spectroscopy for high-fidelity chemical detection. Quantum algorithms for vibrational dynamics are emerging as a promising alternative to resource-demanding classical approaches, but this domain is largely underdeveloped compared to quantum simulations of electronic structure. 
In this work, we describe in detail three distinct forms of the vibrational Hamiltonian: canonical bosonic quantization, real space representation, and the Christiansen second-quantized form. Leveraging Lie algebraic properties of each, we develop efficient fragmentation schemes to enable the use of Trotter product formulas for simulating time evolution. We introduce circuits required to implement time evolution in each form, and highlight factors that contribute to the simulation cost, including the choice of vibrational coordinates. Using a perturbative approach for the Trotter error, we obtain tight estimates of T gate cost for the simulation of time evolution in each form, enabling their quantitative comparison.
Combining tight Trotter error estimates and efficient fragmentation schemes, we find that for the medium-sized CH$_4$ molecule with 9 vibrational modes, time evolution for approximately 1.8 ps may be simulated using as little as 36 qubits and approximately $3 \times 10^{8}$ T gates -- an order-of-magnitude speedup over prior-art algorithms.
Finally, we present calculations of vibrational spectra using each form to demonstrate the fidelity of our algorithms. This work presents a unified and highly optimized framework that makes simulating vibrational dynamics an attractive use case for quantum computers.

\end{abstract}

\maketitle



\section{Introduction}
Quantum mechanical simulations of vibrational dynamics have wide applications in chemistry \cite{puzzarini2019, althorpe2024} and are fundamental for understanding the vibrational structure of molecules \cite{jager2024, nandi2025}, elucidating reaction mechanisms \cite{schroder2022, shi2023}, and obtaining vibrational spectra from first principles  \cite{larsson2022,sunaga2024,larsson2025}. In addition to requiring high-quality potential energy surfaces, these simulations demand either calculating many vibrational eigenstates or explicitly constructing the vibrational Hamiltonian propagator. Numerous algorithms for these calculations on classical computers exist \cite{beck2000,rakhuba2016,baiardi2017,baiardi2019,larsson2019,kallullathil2021,hino2022,kallullathil2023,glaser2023, larsson2024, larsson2025, rano2025}, including sophisticated approximate methods \cite{christoffel1982,fujisaki2007,neff2009,christiansen2004b,jensen2025} (we refer the reader to reviews for details \cite{bowman2008,carrington2017,sibert2019,bowman2022}). However, for large systems, numerically exact quantum mechanical calculations of these quantities are typically not possible due to steep scaling of the computational costs with system size.

With recent developments in quantum technology,  quantum chemistry is emerging as one of the most promising application areas of quantum computing~\cite{kassal2011,bauer2020,mcardle2020,ollitrault2021}. Unlike electronic structure theory, which has enjoyed significant interest from the quantum computing community for a long time, algorithms for \textit{vibrational} spectroscopy and dynamics have only recently begun attracting attention. Most of the focus has been in variational algorithms for near-term hardware \cite{mcardle2019,ollitrault2020,lotstedt2021,lotstedt2022,lee2022,majland2023,nguyen2023,wang2023,asnaashari2024,majland2024,szczepanik2024, knapik2025}, with only a few works focused on fault-tolerant algorithms \cite{sawaya2020,sawaya2021, trenev2025,loaiza2025,kamakari2025,plis2025} --- either without reporting resource requirements or, in most cases, with prohibitively large ones. There is thus a strong need for developing advanced, highly optimized quantum algorithms that would make vibrational dynamics simulations not only feasible, but attractive. 

The choice of the form of the vibrational Hamiltonian is one of the most important factors in determining the cost of simulating vibrational dynamics. Most algorithms focus on either the canonical bosonic second-quantized form that uses vibrational normal modes and bosonic ladder operators $b^\dagger_i, b_i$ to express the Hamiltonian; or the so-called Christiansen second-quantized form \cite{christiansen2004,glaser2023,trenev2025} that unravels the bosonic ladder of states, associating a single particle basis function to each possible bosonic occupancy count, employing hardcore-boson-like creation/annihilation operators $a^\dagger_i, a_i$. In both of these approaches, the occupation number vectors of the relevant basis functions can be straightforwardly encoded into qubits. 

Recently, some of us have proposed an algorithm for vibrational spectroscopy in Ref.~\cite{loaiza2025} that uses the real space form of the Hamiltonian \cite{macridin2018a,macridin2018b}, leveraging methods first introduced in the context of vibronic dynamics~\cite{motlagh_vibronic}. It provides a complementary approach where the amplitudes of the vibrational wavefunction on a real space position grid are encoded into qubits. This real space approach originated in coupled fermion-boson problems \cite{macridin2018a,macridin2018b} and has seen application to variational algorithms for the vibrational problem \cite{lee2022,asnaashari2024}, but it is overall less common, particularly in the context of fault-tolerant quantum algorithms~\cite{loaiza2025,motlagh_vibronic}. Building on this real space algorithm, Ref.~\cite{loaiza2025} provided the first complete, constant-factor resource estimates for a fault-tolerant approach to vibrational spectroscopy. By combining perturbative methods for tightly assessing Trotter error and numerous optimizations of the real space algorithm, the authors achieved the lowest costs reported for a fault tolerant algorithm, significantly improving on prior approaches \cite{sawaya2021,trenev2025}. 

However, despite this recent progress on reducing cost of vibrational Hamiltonian simulation, a full description of the most advanced known approach within each form, as well as a direct comparison of their relative strengths and weaknesses, is currently lacking.

In this work, we provide the first comprehensive description of how to perform time evolution with all known vibrational Hamiltonian forms. In doing so, we introduce additional techniques and optimizations that together yield a tenfold speedup relative to previous state-of-the-art methods for vibrational Hamiltonian simulation \cite{loaiza2025}. We focus on Trotter product formulas for time evolution because their low ancilla overhead makes them highly suitable for early fault-tolerant algorithms compared to qubitization-based approaches -- and make three key contributions.

First, for all three Hamiltonian forms, we introduce schemes to partition vibrational Hamiltonians into efficient fast-forwardable fragments that allow for low-cost implementation of Trotterized time evolution. These are based on a generalization of the \gls{CSA} approach \cite{yen2021} that exploits Lie algebraic properties of the Hamiltonian to perform fragmentation. Moreover, for all these cases, we show how to implement Trotterized time evolution for those fragment decompositions by providing explicit quantum circuit constructions. 

Second, we carefully derive and compare the qubit and T gate costs of implementing time evolution for all Hamiltonian forms and fragmentation schemes presented. Crucially, we provide \textit{tight} estimates of Trotter error, which affects the overall cost of time evolution by controlling the number of Trotter time steps required for a particular fragmentation scheme to stay below an overall target simulation error. To do this, we employ a recently introduced perturbative method \cite{Babbush2015,mehendale2025,loaiza2025}: we find this approach provides much tighter error estimates than previous upper bounds based on nested fragment commutators \cite{childs2021}, which are known to be loose and poorly correlated with the error in eigenenergy calculations \cite{trenev2025,mehendale2025}. 

Third, we combine these efficient Hamiltonian fragmentations and tight Trotter error estimates with other optimizations of time evolution implementation, such as the use of local-mode coordinates \cite{majland2023}, and many of the optimizations introduced in Ref.\cite{loaiza2025}, to achieve a factor of 10 speedup for time evolution of vibrational Hamiltonians over prior-art algorithms \cite{loaiza2025}. To demonstrate this, we perform constant-factor resource estimates on a few select molecules for implementing time evolution for a given total evolution time and fixed accuracy for the three forms and fragmentations, and highlight strengths and weaknesses of each approach. As an example, we find that for the medium-sized CH$_4$ molecule with 9 vibrational modes, time evolution for approximately 1.8 ps may be simulated using 36 qubits and approximately $3 \times 10^{8}$ T gates using an optimized Christiansen fragmentation approach --  representing a tenfold speedup compared to the real space algorithm in Ref.~\cite{loaiza2025}.

Finally, as a demonstration of the fidelity of our time evolution implementation and Trotter error estimation for all fragmentation schemes and Hamiltonian forms, we use the open-source library PennyLane \cite{pennylane} and its simulator backend \texttt{lightning} to simulate the time evolution circuits as part of an algorithm to obtain vibrational spectra \cite{loaiza2025}, confirming correctness against a classical reference spectrum.

The rest of the article is arranged as follows. Section~\ref{sec:theory} describes the construction of the vibrational Hamiltonian, presents the unified Lie algebraic perspective on fragmentation, and outlines the perturbative approach used to obtain estimates of the Trotter time step. Then, in sections~\ref{sec:c-form}-\ref{sec:bosonic}, we derive a fragmentation scheme for each Hamiltonian form, demonstrate how to perform Trotterized time evolution, and perform constant-factor resource estimation. Section~\ref{sec:results} presents resource estimates and spectra for the various methods and Sec.~\ref{sec:comp_forms} includes a discussion of their strengths and weaknesses. Section~\ref{sec:conc} presents our conclusions. 

\section{Theory} \label{sec:theory}

\subsection{Vibrational Problem} \label{sec:setup}

We begin by describing how one constructs a vibrational Hamiltonian for a molecule of interest. Focusing on systems that do not access regions of nonadiabatic coupling, such that the Born-Oppenheimer approximation remains valid, the construction of the vibrational Hamiltonian assumes that the system stays in the electronic ground state. The vibrational \gls{PES} is then defined as the energy of the ground electronic state as a function of nuclear geometry $ V_N(\textbf{R})$ for $\textbf{R}=\{\vec R_1,\cdots,\vec R_{L}\}$ the Cartesian coordinates of all $L$ nuclei making up the molecular system. The nuclear Hamiltonian under the Born-Oppenheimer approximation then corresponds to
\begin{equation}
     H_{\rm nuc} =  T_N+ V_N(\textbf{R}),
\end{equation}
where we have defined the nuclear kinetic energy operator
\begin{equation}
     T_N = -\sum_{\alpha=1}^{L} \frac{1}{2m_\alpha} \nabla^2_\alpha
\end{equation}
for $m_\alpha$ the mass of nucleus $\alpha$. Here and in the remainder of this work we use atomic units, unless stated otherwise. Note that even though we are only considering the electronic ground state, it has been shown that including terms from the diagonal Born-Oppenheimer correction can improve the quality of the \gls{PES} \cite{dboc_1,dboc_2}.  The addition of these terms entails a modification of $V_N(\textbf{R})$, which does not affect our discussion. Molecular vibrations occur around an equilibrium configuration that corresponds to a local minimum in the \gls{PES} with associated geometry $\textbf{R}^{({\rm eq})}$. This geometry being a minimum implies that all terms of the Jacobian in the equilibrium geometry vanish, namely 
\begin{equation} \label{eq:eq_jacobian}
    \frac{\partial  V_N}{\partial R_{\alpha\rho}}\Bigg|_{\textbf{R}^{({\rm eq})}} = 0
\end{equation}
for $\rho=x,y,z$ a Cartesian direction. Normal, \textit{mass-weighted} coordinates $Q_i$ are then obtained by diagonalizing the Hessian matrix of the PES in this equilibrium geometry, from which we can define a transformation between Cartesian and normal coordinates as
\begin{equation}
    R_{\alpha\rho} = R_{\alpha\rho}^{({\rm eq})} + \sum_{i=1}^M B_{\alpha\rho,i} Q_i,
\end{equation}
where we have defined the number of vibrational modes as $M=3L-6$ (or $3L-5$ for linear molecules), and $B_{\alpha\rho,i}$ as an element from the $3L\times M$-dimensional matrix which defines the $M$ displacement vectors associated to each normal mode. $B_{\alpha\rho,i}$ is obtained by diagonalizing the Hessian in mass-weighted coordinates. The equilibrium geometry then corresponds to $\vec Q=0$. Note that the remaining $6$ (or $5$) degrees of freedom not accounted for correspond to the translational and rotational coordinates of the centre of mass of the molecule. In this work we neglect the centre of mass motion of the molecule, as well as the rotational and rovibrational contributions to the Hamiltonian (which are accounted for in the more complex Watson Hamiltonian \cite{watson1968simplification,armino2020computational,gong2018fourth}), while noting that the techniques proposed in this work can be straightforwardly extended to the case where these contributions are also considered. 

In the normal-mode coordinates, the Hessian becomes diagonal, with eigenvalues being the associated frequencies $\omega_i$. Defining the normal coordinates in natural units as
\begin{equation}
    q_i = \sqrt{\omega_i} Q_i
\end{equation}
with the associated momentum in natural units $p_j=-i\partial_{q_j}$ transforms the vibrational component of kinetic energy operator into the form
\begin{equation} \label{eq:kin_normal}
    T_{\rm vib} = \frac{1}{2} \sum_{i=1}^M \omega_i p_i^2,
\end{equation}
and the vibrational Hamiltonian can be written as, 
\begin{align}
    H = T_{\rm vib} + V(\bm{q}). \label{eq:h_vib}
\end{align}

Rather than working with normal modes, in practice it is sometimes preferable to introduce the concept of mode localization, as discussed in Refs.~\cite{mode_loc_1,mode_loc_2,mode_loc_3,mode_loc_4}. In analogy with orbital localization techniques in electronic structure, the idea behind mode localization is to define a unitary transformation $\textbf{U}$ of normal modes with associated matrix elements $U_{ij}$ as to obtain a localized set of vibrational modes $\tilde q_i$:
\begin{equation}
    \tilde q_i = \sum_{j=1}^M U_{ij} q_i.
\end{equation}
The $M\times M$ unitary matrix $\textbf{U}$ is obtained by maximizing the atomic character of each particular vibrational mode. The localized vibrational modes obtained after the procedure will have displacement vectors that have been optimized to each involve as few atoms as possible, while remaining orthogonal. This is completely equivalent to the Pipek-Mezey orbital localization technique used in electronic structure \cite{pipek_mezey} (or Wannierization as it is sometimes known in periodic systems), which corresponds to the maximization problem,
\begin{equation}
    \textbf{U}^*= \argmax_{\textbf{U}} \left\{ \sum_{j=1}^M \sum_{\alpha=1}^{L} \left( \sum_{\rho=x,y,z} \left( \sum_{i=1}^M U_{ij} \tilde B_{\alpha\rho,i} \right)^2 \right)^2 \right\},
\end{equation}
where we have defined the normalized displacement vectors,
\begin{equation}
    \tilde B_{\alpha\rho,i} = \frac{B_{\alpha\rho,i}}{\sqrt{\sum_{\tilde\alpha\tilde\rho} |B_{\tilde\alpha\tilde\rho,i}|^2}}.
\end{equation}
Note that the localized modes do not diagonalize the Hessian matrix of the \gls{PES}, and the resulting kinetic energy operator is also no longer diagonal in these coordinates:
\begin{equation} \label{eq:final_kinetic}
    T_{\rm vib} = \sum_{i,j=1}^M \tilde p_i\tilde p_j \sum_{k=1}^M \frac{\omega_k}{2} U_{ki} U_{kj}.
\end{equation}
Regardless of whether we chose to work with the canonical normal modes or with their localized counterparts, the vibrational Hamiltonians take the form shown in Eq.~\eqref{eq:h_vib}. Here and for the remainder of this work we will simply use the notation of $q_i$'s to refer to the vibrational modes, noting that both choices yield the same functional form for the \gls{PES}.

We now unpack the $M$-variable \gls{PES} $V(\bm{q})$ by writing it as an $n$-mode expansion \cite{Carter1997,bowman2003,Toffoli2007}
\begin{align}
    V(\bm{q}) & =  \mathcal{V}^{(0)} + \sum_{i=1}^{M} \mathcal{V}^{(1)}_i(q_i) + \sum_{i>j}^{M} \mathcal{V}^{(2)}_{ij}(q_i, q_j) \notag \\ 
    & + \sum_{i>j>k}^{M} \mathcal{V}^{(3)}_{ijk}(q_i, q_j, q_k) +\cdots, \label{eq:n_mode}
\end{align}
where each of the terms of the expansion corresponds to
\begin{align}
    \mathcal{V}^{(0)} &= V(\bm{q}=0), \\
    \mathcal{V}_{i}^{(1)}(q_i) &= V(0,\cdots,0,q_i,0,\cdots,0) - \mathcal{V}^{(0)}, \\
    \mathcal{V}_{ij}^{(2)}(q_i,q_j) &= V(0,\cdots,q_i,\cdots,q_j,\cdots,0) \nonumber \\
    &\ \ \ \ -\mathcal{V}_{i}^{(1)}(q_i) -\mathcal{V}_{j}^{(1)}(q_j)-\mathcal{V}^{(0)}, 
\end{align}
and so on, where $\mathcal{V}^{(n)}$ depends only on $n$ of the $M$ modes which are denoted by the subscript. Each $\mathcal{V}^{(n)}$ is generally an anharmonic (i.e. non-quadratic) function of the $q_i$'s: the expansion becomes exact if expanded up to the $M$th order. 

The advantage of using this expansion comes from the fact that the properties from the full-dimensional \gls{PES} can be generally obtained by truncating the $n$-mode expansion to a low $n$, namely setting $\mathcal{V}^{(m)}$'s to $0$ for all $m>n$ \cite{Carter1997,bowman2003,Toffoli2007}. In addition, it has been shown that local-mode Hamiltonians generally allow one to use a lower $n$-mode truncation than normal-mode Hamiltonians to achieve equivalent accuracy. For example, for the ethane $\rm C_2H_4$ molecule, a $2$-mode expansion using localized modes was able to recover the same accuracy as a $4$-mode expansion using normal modes \cite{mode_loc_1}. As will be shown in upcoming sections, the algorithmic cost of simulating the vibrational system in both Christiansen and real space forms depends strongly on the degree in the $n$-mode expansion, highlighting how significant the mode localization technique can be in reducing simulation costs. 

While the Christiansen form requires the use of the $n$-mode expansion of the potential $V(\bm{q})$, the bosonic and real space forms generally rely on the Taylor expansion of the potential $V(\bm{q})$ to a given order $d$ around the equilibrium geometry. To allow direct comparison of the constant factor algorithmic cost of these forms, in this work, we use the $n$-mode expansion of the potential throughout, and further approximate each $\mathcal{V}^{(i)}$ to fourth order in a Taylor expansion. For example, we use the $2$-mode expansion along with a $4$th-order Taylor expansion and call the Hamiltonians 2M4T Hamiltonians. Such a 2M4T Hamiltonian can be obtained in both normal and local modes. 

Lastly, we mention a couple of choices that can be made for the Christiansen form that yield an improved description of the vibrational problem at no additional simulation cost. Firstly, the use of a Taylor expansion of each $\mathcal{V}^{(i)}$ is not required for the Christiansen form, but has only been considered here to facilitate a fair comparison between the different forms. Thus, we also consider $2$-mode Hamiltonians in the Christiansen form without invoking a Taylor approximation. We call these 2M$\infty$T Hamiltonians. Furthermore, unlike the bosonic form, which uses the \gls{HO} basis for each vibrational mode, the Christiansen approach allows greater flexibility in the choice of basis functions, also referred to as \textit{modals}. Thus, we also include cost estimates with modals obtained from a \gls{VSCF} calculation  \cite{vscf_1,vscf_2,vscf_3,vscf_4} (details in Sec.~\ref{sec:vscf}), which account for anharmonicity in each mode at a mean-field level, and thus provide a better description than the same number of \gls{HO} modals per mode.

\subsection{Unified Perspective on Hamiltonian Fragmentation} \label{sec:lie_algebra}

\textbf{Motivation for Fragmentation:} Trotter product formulas rely on decomposing the Hamiltonian into \textit{fast-forwardable} fragments, 
\begin{align}\label{eq:part}
    H = \sum_{\nu} H_{\nu}.
\end{align} 
These fast-forwardable fragments $H_{\nu}$ can be diagonalized using known unitary operators $\mathcal{U}_{\nu}$, as, 
\begin{align}\label{eq:Hnu}
    H_{\nu} = \mathcal{U}_{\nu}D_{\nu}\mathcal{U}_{\nu}^{\dagger},
\end{align}
where $D_{\nu}$ is the diagonalized fragment. As a consequence, each fragment can be easily exponentiated to obtain its propagator, 
\begin{align} 
    e^{-iH_{\nu}t} = \mathcal{U}_{\nu}e^{-iD_{\nu}t}\mathcal{U}_{\nu}^{\dagger}. \label{eq:cartan_evolution}
\end{align}
The propagators of individual fragments are then used to construct the approximate, Trotterized propagator, as mentioned in Sec.~\ref{sec:trotter_error}. As a side note, we would like to mention that for many quantum algorithms, such as those based on \gls{QPE}, a controlled implementation of the time evolution is required. The form of each fragment propagator in Eq.~\eqref{eq:cartan_evolution} also allows for an efficient controlled implementation, as will be discussed briefly in Sec.~\ref{sec:cf_te}. \\

\textbf{Fragmentation Framework:} 
Although the decomposition of the Hamiltonian into fast-forwardable fragments is necessary to implement the propagator using Trotter product formulas, finding a decomposition that leads to an efficient implementation is a non-trivial task. Here, we introduce a unified framework for fragmenting the Hamiltonian based on generalization of the \gls{CSA} approach \cite{yen2021}. The Hamiltonian $H$ can always be written as a polynomial of elementary operators $\{A_k\}$, 
\begin{align}
    H = \sum_k c_k A_k + \sum_{k,l} d_{k,l} A_k A_l + ...
\end{align}
The operators $\{A_k\}$ are generators of a Lie algebra $\mathcal{A}$ with respect to the commutator,
\begin{align}
    & \left[A_j , A_k \right] = \sum_l \alpha_{l}^{(j,k)} A_l, 
\end{align}
and $\alpha_{l}^{(j,k)}$ are the structure constants. In general, the products $A_k A_l$ and higher-order products do not belong to the Lie algebra $\mathcal{A}$. The Hamiltonian is thus an element of the universal enveloping algebra \cite{yen2021}. For many problems in physics and chemistry, the Lie algebra $\mathcal{A}$ is known and well studied. 
In upcoming sections we will look at different realizations of the vibrational Hamiltonian using different Lie algebras $\mathcal{A}$. 

To find fragments of the Hamiltonian, we exploit the maximum abelian subalgebra (MASA) with elements representable by normal operators $\mathcal{C} \subset \mathcal{A}$, i.e. the elements of the MASA $C_k$, commute with each other,
\begin{align}
   & \left[C_k, C_l \right] = 0, & C_k, C_l \in \mathcal{C}
\end{align}
and can be diagonalized. 
Since the MASA elements $C_k$ commute, polynomials of MASA elements 
\begin{align}
    p(\{C_k\}) = \sum_k \lambda_k C_k + \sum_{k,l} \eta_{k,l} C_k C_l + ...
\end{align}
also form a linear combination of commuting elements in the universal enveloping Lie algebra.
In all used Lie algebras there is a representation of MASA elements as diagonal matrices. Thus, $D_\nu$ can be always taken as a polynomial of MASA elements. 

All finite Lie algebras are classified and their MASAs with elements that are representable as normal operators are known as well. 
For semi-simple Lie algebras (e.g. $\mathfrak{su}(N)$) such MASAs
are also the Cartan subalgebras~\cite{hall,gilmore2008}. Previously, the fragmentation approach presented in Ref.~\cite{yen2021} considered Lie algebras that were semi-simple, 
thus all MASAs were CSAs. 
In this work, not all considered Lie algebras are semi-simple, for example the Heisenberg Lie algebra, $\{1,x,p\}$, is nilpotent. 
For non-semi-simple Lie algebras, CSAs are not necessarily MASAs with 
operators that are representable with normal operators. Thus, we will need to go beyond the CSA approach of Ref.~\cite{yen2021} in these cases.  

Conjugating MASAs representable as normal operators by unitary transformations, 
\begin{align}
    C^{(\nu)}_k = \mathcal{U}_{\nu} C_k \mathcal{U}_{\nu}^\dagger
\end{align}
creates isomorphic commuting operators $C^{(\nu)}_k$ which contain generators outside of the original MASA. In the case of semi-simple Lie algebras in the CSA approach of Ref.~\cite{yen2021}, $\mathcal{U}_{\nu}$ were 
chosen to be elements of the corresponding Lie group. 
In this work some of the Lie algebras are not semi-simple, and thus $\mathcal{U}_{\nu}$ need to be different than 
elements of the Lie group. Yet, in all cases the fragments 
are defined as polynomials of unitarily transformed MASA elements,
\begin{align}
    H_\nu = \sum_k \lambda^{(\nu)}_k C^{(\nu)}_k + \sum_{k,l} \eta^{(\nu)}_{k,l} C^{(\nu)}_k C^{(\nu)}_l + ... \label{eq:sol_frag}
\end{align}
To obtain the partitioning of the Hamiltonian as in Eq.~\eqref{eq:part} the fragment parameters $\{\lambda^{(\nu)}_k\}$, $\{\eta^{(\nu)}_k\}$ and $\mathcal{U}_{\nu}$ are optimized so that $H_\nu$'s satisfy \eq{eq:part}.      
\subsection{Trotter Error Analysis} \label{sec:trotter_error}
The Hamiltonian fragments, $H = \sum_{\nu} H_{\nu}$, will be used in conjunction with the Trotter product formula to approximate the propagator $e^{-iH\tau}$. A simulation time step of length $\tau$ is generally divided into $r$ time steps of time $\Delta t = \tau/r$, which is then approximated using a product formula. In this work, we use the second-order Trotter formula,
\begin{align}
    e^{-iH \Delta t} \approx \prod_{\nu=1}^{N_f} e^{-iH_{\nu}\Delta t/2} \prod_{\nu=N_f}^{1} e^{-iH_{\nu}\Delta t/2} \equiv e^{-iH_{\text{eff}}(\Delta t)\Delta t}. \label{eq:trot2} 
\end{align}
Although the use of higher-order product formulas can allow for larger Trotter time steps $\Delta t$, the circuit depth increases exponentially with the order of the approximation \cite{berry2007}. That said, the following analysis can nevertheless be generalized to higher-order product formulas. 

To keep the error arising from the Trotter approximation within some acceptable error, the Trotter time step $\Delta t$ is generally chosen using theoretical upper bounds on the Trotter error.
Frequently used commutator bounds \cite{childs2021,reiher2017} are known to be loose and have recently been found to be poorly correlated with error in eigenenergy calculations \cite{mehendale2025}. Here we instead use Trotter error estimates based on perturbation theory \cite{mehendale2025,loaiza2025} to choose the Trotter time step $\Delta t$. Equation~\eqref{eq:trot2} indicates that approximate propagation obtained from a Trotter product formula can instead be viewed as propagation under an approximate (or effective) Hamiltonian $H_{\text{eff}}(\Delta t)$. Thus, the error in the eigenvalues obtained from QPE when employing the Trotter product formula is due to the difference between the eigenvalues of $H$ and $H_{\text{eff}}(\Delta t)$. Following Refs.~\cite{mehendale2025,loaiza2025}, this error can be estimated using perturbation theory as follows. The effective Hamiltonian, $H_{\text{eff}}(\Delta t)$, can be obtained using the Baker-Campbell-Hausdorff formula as, 
\begin{align}
    H_{\text{eff}}(\Delta t) = H + \sum_k \Theta_k \Delta t^k,
\end{align}
and $\Theta_k$ involve nested commutators of the fragments. For the second-order Trotter formula used here, we have, $\Theta_1 = 0$, and 
\begin{align}
    \Theta_2  & = \frac{1}{12}\sum_{\nu=0}^{N_f}\sum_{\mu,\xi < \nu} \left[\left[H_{\mu},H_{\nu}\right],H_{\xi}\right] \notag \\
    & + \frac{1}{24}\sum_{\nu=0}^{N_f}\sum_{\mu< \nu} \left[\left[H_{\mu},H_{\nu}\right],H_{\nu}\right]. \label{eq:V2}
\end{align}
The $n$th eigenvalue of $H_{\text{eff}}(\Delta t)$ can then be calculated using first-order time-independent perturbation theory as,
\begin{align}
    E^{(n)}_{\text{eff}} = E^{(n)} + \epsilon^{(n)}_2 \Delta t^2 + \mathcal{O}\left(\Delta t^3\right),
\end{align}
where,
\begin{align}
    \epsilon^{(n)}_2 = \mel{\Psi_n}{\Theta_2}{\Psi_n}.
\end{align}
Note that $\ket{\Psi_n}$, the exact eigenstates of $H$ are required to obtain the effective eigenenergies. However, tests on both electronic \cite{mehendale2025} and vibrational \cite{loaiza2025} Hamiltonians for several molecules have demonstrated that error estimates obtained using mean-field eigenstates are well correlated with the exact error. Here, we employ either \gls{HO} or \gls{VSCF} eigenstates, denoted $\ket{{\Phi}_n}$, to approximate $\epsilon^{(n)}_2$ as,
\begin{align}
    \tilde{\epsilon}^{(n)}_2 = \mel{{\Phi}_n}{\Theta_2}{{\Phi}_n}. \label{eq:nth_trot_error}
\end{align}

The perturbative error estimate for the $n$th state is then defined as,
\begin{align}
    \Delta E^{(n)} = \left |E^{(n)}_{\text{eff}} - E^{(n)} \right| \approx  \left| \tilde{\epsilon}^{(n)}_2 \right| \Delta t^2.
\end{align}

We also note that the inclusion of higher-order terms in the expansion of $H_{\text{eff}}(\Delta t)$, or the use of higher-order perturbation theory will both result in third-order and higher contributions to the energies. Given that the Trotter time step $\Delta t$ is generally small, these higher-order contributions can generally be neglected to leading order. We are interested in the Trotter error for some initial state $\ket{\Psi} = c_n \ket{{\Phi}_n}$, which in similar spirit is approximated by, 
\begin{align}
    \Delta E \approx  \left(\sum_n |c_n|^2 \left|\tilde{\epsilon}^{(n)}_2\right|\right) \Delta t^2 = \tilde{\epsilon}_2 \Delta t^2, \label{eq:trial_state_trot_error}
\end{align}
where the last equality defines $\tilde{\epsilon}_2$. The Trotter time step length is then chosen as $\Delta t = \sqrt{\epsilon_{\text{trot}}/\tilde{\epsilon}_2}$. In this work we conservatively chose $\epsilon_{\text{trot}} = 7  \text{ cm}^{-1}$ as the error allowed; this heuristic was the largest value for which the associated absorption spectra remained visually indistinguishable from exact simulations. The number of Trotter time steps $r$ is then defined as,
\begin{align}
    r = \left\lceil \frac{\tau}{\Delta t} \right\rceil,
\end{align}
where $\lceil . \rceil$ denotes the ceiling function.

\section{Christiansen form} \label{sec:c-form}

\subsection{Hamiltonian}

The Christiansen formalism is a second-quantized formalism for representing the vibrational Hamiltonian \cite{christiansen2004}, fundamentally based on the $n$-mode expansion of the \gls{PES}, Eq.~\eqref{eq:n_mode}. In a system with $M$ vibrational modes, for the $l$th vibrational mode the Christiansen formalism uses $N_l$ single-mode basis functions $\phi^{(l)}_{\alpha_l}(q_l)$ for $\alpha_l=0,\dotsi,N_l-1$, referred to as \textit{modals}. A simple choice for the modals is to use eigenfunctions of the harmonic oscillator, corresponding to different number of bosons in the vibrational mode: however, other basis choices are also possible. These single-mode basis functions are then used to form a direct product basis $\Phi(\bm{q}) = \prod_{l=1}^{M} \phi^{(l)}_{\alpha_l}(q_l)$, which corresponds to the $\alpha_l$th modal being occupied for the $l$th mode. These basis functions are represented by \glspl{ONV},
\begin{multline}
\ket{\bm{k}}  = \left| k_0^{(1)},\dotsi,k^{(1)}_{N_1-1},       \dotsi,k^{(l)}_0,\dotsi,k^{(l)}_{N_l-1}, \right. \\
\dotsi,k^{(M)}_0,\dotsi,k^{(M)}_{N_M -1}\rangle, \notag
\end{multline}
with $k^{(l)}_{\alpha_l} \in \{0,1\}$ being the occupation number for the $\alpha_l$th modal for the $l$th mode. Since only one modal from each mode is present in the direct product basis function $\Phi(\bm{q})$, because there can only be one concrete number of bosons occupying the vibrational mode in a given product state, physically meaningful \glspl{ONV} form a subspace of the full \gls{ONV} space, satisfying the condition, 
\begin{align}
    \sum_{\alpha_l = 0}^{N_l-1} k^{(l)}_{\alpha_l} = 1, \quad l = 1,\dots, M. \label{eq:sepm_cond}
\end{align}
We can define creation and annihilation operators that act on these \glspl{ONV} as,
\begin{align}
    a_{\alpha_l}^{\dagger} \ket{\bm{k}} & = \ket{\bm{k} + \bm{1}^{(l)}_{\alpha_l}} \\
    a_{\alpha_l} \ket{\bm{k}} & = \delta_{0,\alpha_l}\ket{\bm{k} - \bm{1}^{(l)}_{\alpha_l}},
\end{align}
where $\ket{\bm{1}^{(l)}_{\alpha_l}}$ is an \gls{ONV} with  $n^{(l)}_{\alpha_l}=1$ and all other entries zero. They follow commutation relations,
\begin{align}
    \left[a_{\alpha_l},a_{\beta_m}^{\dagger}\right] &= \delta_{l,m}\delta_{\alpha,\beta}, \\  \left[a_{\alpha_l},a_{\beta_m}\right] & = \left[a_{\alpha_l}^{\dagger},a_{\beta_m}^{\dagger}\right] = 0.
\end{align}
A useful intuition for these operators is that they can be seen as a ``factorization'' of the regular harmonic oscillator ladder operators
\begin{equation}
    b^\dagger_l \to \sum_{\alpha_l} \sqrt{\alpha_l + 1} a^\dagger_{\alpha_{l}+1} a_{\alpha_l}.
\end{equation}
Indeed, to add one more boson $b^\dagger_l$ to a vibrational mode, in this representation one needs to remove the occupancy from the $\alpha_l$'th modal and occupy the $\alpha_{l+1}$'th modal: since the state could have any boson occupancy, it is necessary to sum over the modals $\alpha_l$ to capture the possibility that $b^\dagger_l$ adds a boson to any given occupation.

The vibrational Hamiltonian from Eq.~\eqref{eq:h_vib} in this representation, to which we refer as the Christiansen form, is written as \cite{christiansen2004},
\begin{align}
    H &= \sum_{l=1}^{M}\sum_{\alpha_l \beta_l = 0}^{N_l-1} h^{(l)}_{\alpha_l \beta_l} E^{\alpha_l}_{\beta_l} \notag \\
     & + \sum_{l > m =1}^{M} \sum_{\alpha_l \beta_l = 0}^{N_l-1} \sum_{\gamma_m \delta_m = 0}^{N_m-1} g^{(lm)}_{\alpha_l \beta_l \gamma_m \delta_m} E^{\alpha_l}_{\beta_l} E^{\gamma_m}_{\delta_m} \label{eq:c_form} \\
     & \equiv H_{1M} + H_{2M}.
\end{align}
Here we have restricted the Hamiltonian to at most two-mode couplings for the sake of brevity. However, higher mode couplings can also be included, and the rest of the following discussion can be generalized for those cases. 

The indices $l,m$ go over the $M$ vibrational modes of the molecule. Indices of the form $\alpha_l$ correspond to the $\alpha$th modal of the $l$th mode, with a total of $N_l$ modals for the $l$th mode. From here on, we do not explicitly write ranges for summations for the sake of brevity. The operators 
\begin{align}
    E^{\alpha_l}_{\beta_l} \equiv a_{\alpha_l}^{\dagger}a_{\beta_l} \label{eq:exc_ops}
\end{align}
are often called excitation operators. The coefficients of the Hamiltonian in this form can be obtained through the following integrals of the kinetic energy and $n$-mode expansion of the \gls{PES} in Eqs.~\eqref{eq:final_kinetic} and \eqref{eq:n_mode} as,
\begin{align}
    h_{\alpha_l\beta_l}^{(l)} &= \int_{-\infty}^\infty dq_l \, \phi^{(l)}_{\alpha_l}(q_l) \left(T_{\rm vib} + \mathcal{V}^{(1)}_l(q_l)\right) \phi^{(l)}_{\beta_l}(q_l),
\end{align}
\begin{align}
    g_{\alpha_l\beta_l\gamma_m\delta_m}^{(lm)} &= \int_{-\infty}^\infty dq_l \int_{-\infty}^\infty \, dq_m \phi^{(l)}_{\alpha_l}(q_l) \phi^{(m)}_{\gamma_m}(q_m)  \nonumber \\
    &\ \ \ \  \times \mathcal{V}^{(2)}_{lm}(q_l,q_m) \phi^{(l)}_{\beta_l}(q_l) \phi^{(m)}_{\delta_m}(q_m).
\end{align}
Note that in the case where the vibrational modes are localized, the $T_{\rm vib}$ kinetic energy operator becomes non-diagonal, which effectively makes it a two-mode operator that needs to also be included in the two-mode integral. In practice these integrals can be done efficiently by using a Gauss-Hermite quadrature and \gls{HO} basis functions over each vibrational mode. In addition, the kinetic components can be obtained analytically by expressing the momentum operators with bosonic ladder operators and directly applying them on the \gls{HO} basis functions.

We note in passing that since physically meaningful \glspl{ONV} form a subspace satisfying the condition defined in Eq.~\eqref{eq:sepm_cond}, the Hamiltonian has $M$ distinct symmetries,
\begin{align}
& S_l =  \sum_{\alpha_l = 0}^{N_l-1} n_{\alpha_l}, & l = 1,\dotsi,M &
\end{align}
satisfying $[H,S_l] = 0$. Here $n_{\alpha_l} = E^{\alpha_l}_{\alpha_l} = a_{\alpha_l}^{\dagger}a_{\alpha_l}$ are number operators. These symmetries could be exploited to reduce the number of Trotter steps for a more efficient Trotter based simulation \cite{tran2021}. Furthermore, they could also be used to reduce the one-norm of the Hamiltonian \cite{loaiza2023a,loaiza2024b}, which is one of the factors determining the cost of quantum algorithms based on the linear combination of unitaries method. However, these improvements are beyond the scope of this work.

\subsection{VSCF} \label{sec:vscf}
Once the Christiansen form of the Hamiltonian has been obtained [Eq.~\eqref{eq:c_form}], the \gls{VSCF} procedure can be used to refine the modal basis \cite{vscf_1,vscf_2,vscf_3,vscf_4}. Note that the \gls{VSCF} procedure can also be done for bosonic or real space forms of the Hamiltonian, although for simplicity we only discuss it here for the Christiansen form. The idea is to use modal rotations over a particular mode $l$, allowing us to define a rotated basis for mode $l$ as
\begin{equation} \label{eq:modal_rotation}
    \ket{\tilde{\phi}_{\alpha_l}^{(l)}(\kappa^{(l)})} = \sum_{\beta_l} U^{(l)}_{\alpha_l\beta_l}(\kappa^{(l)}) \ket{\phi_{\beta_l}^{(l)}},
\end{equation}
where each $U^{(l)}_{\alpha_l\beta_l}$ is an element from a (real) orthogonal matrix that is parametrized by $\kappa^{(l)}$. Refer to Sec.~\ref{sec:cform_frag} for more details on modal rotations. 
We can then define a one-mode Fock operator as
\begin{equation} \label{eq:fock}
    F^{(l)} = \bra{\prod_{m\neq l}\tilde{\phi}_0^{(m)}(\kappa^{(m)})}  H \ket{\prod_{m\neq l}\tilde{\phi}_0^{(m)}(\kappa^{(m)})},
\end{equation}
with the Fock ground-state energy of the Hamiltonian being defined as
\begin{equation} \label{eq:fock_energy}
    E_0^{(\textrm{Fock})} = \sum_{l=1}^M \bra{\tilde{\phi}_0^{(l)}(\kappa^{(l)})}  F^{(l)} \ket{\tilde{\phi}_0^{(l)}(\kappa^{(l)})}.
\end{equation}
The \gls{VSCF} procedure then corresponds to performing the following steps:
\begin{enumerate}
    \item Start with the Christiansen form of the vibrational Hamiltonian in Eq.~\eqref{eq:c_form}. Initialize the matrices $\kappa^{(l)}$ as zeros with dimensions $N_l\times N_l$ for all modes $l=1,\cdots,M$. Calculate the initial Fock energy as shown in Eq.~\eqref{eq:fock_energy}.
    \item Do a \gls{VSCF} cycle, which corresponds to:
    \begin{enumerate}
        \item Set $l=1$.
        \item Obtain the Fock operator $ F^{(l)}$ [Eq.~\eqref{eq:fock}]. Note that this is a one-mode operator, which can be represented as a Hermitian $N_l\times N_l$ matrix.
        \item Diagonalize $ F^{(l)}$, obtaining the diagonalizing unitary $U^{(l)}(\kappa^{(l)})$. Update the modals for mode $l$ according to Eq.~\eqref{eq:modal_rotation}.
        \item Set $l = l+1$. If $l>M$, the cycle is finished, otherwise return to $2.b$.
    \end{enumerate}
    \item Calculate the Fock energy with the updated modal basis. If the energy change with respect to last calculated energy is smaller than some threshold, then the \gls{VSCF} procedure is finished. Otherwise, return to 2.
\end{enumerate}
Note that some optimizations can be done to reduce the cost of calculating the Fock matrices, which constitute the main computational bottleneck of this procedure. A more in-depth discussion of how this is done is shown in Ref.~\cite{how_to_vscf}. Once the \gls{VSCF} procedure has been finished, we recover a Christiansen Hamiltonian in the same form as in Eq.~\eqref{eq:c_form}, written in the basis of the optimized \gls{VSCF} modals $\ket{\tilde{\phi}^{(l)}_{\alpha_l}}$ (explicit dependence on $\kappa^{(l)}$ omitted here) instead of the \gls{HO} basis functions $\ket{\phi^{(l)}_{\alpha_l}}$.

\subsection{Fragmentation} \label{sec:cform_frag}

Here we discuss two approaches to fragmentation that differ in the Lie algebras used for realization of the Hamiltonian: the first approach partitions the Christiansen Hamiltonian directly, the second converts it to Pauli products first and partitions after. We refer to the former as the \gls{CGF} scheme. \\

\textbf{Christiansen Greedy Fragmentation:} Noting that real Hermitian linear combinations of $E^{\alpha_l}_{\beta_l}$ (Eq.~\eqref{eq:exc_ops}) are generators of the compact Lie algebra $\mathfrak{o}(N_l)$ (or $\mathfrak{so}(N_l)$), the Christiansen Hamiltonian is an element of the universal enveloping algebra of the compact semi-simple Lie algebra $\oplus_{l=1}^{M}\mathfrak{o}(N_l)$. The \gls{CSA} is formed by the number operators $\{n_{\alpha_l} \equiv E^{\alpha_l}_{\alpha_l} \}$. The maximal tori theorem guarantees that the one-mode Hamiltonian $H_{1M}$ can be diagonalized using a unitary from the corresponding Lie group, $\mathcal{G}$, as \cite{hall, yen2021} 
\begin{align}
    \mathcal{G}^{\dagger} \left(\sum_{l} \sum_{\alpha_l \beta_l}  h^{(l)}_{\alpha_l \beta_l} E^{\alpha_l}_{\beta_l}\right) {\mathcal{G}} = \sum_{l} \sum_{\alpha_l} \epsilon^{(l)}_{\alpha_l} n_{\alpha_l}, \label{eq:cform_1m_frag}
\end{align}
where $\mathcal{G} \equiv \prod_{l} \mathcal{G}^{(l)}$, and $\mathcal{G}^{(l)} = e^{\sum_{\alpha_l>\beta_l}\theta_{\alpha_l \beta_l}\left(E^{\alpha_l}_{\beta_l} - E^{\beta_l}_{\alpha_l}\right)}$. The unitary $\mathcal{G}^{(l)}$ mixes creation operators of the $l$th mode as $\mathcal{G}^{(l)} a_{\alpha_l}^{\dagger} {\mathcal{G}^{(l)}}^{\dagger} = \sum_{\beta_l} U^{(l)}_{\alpha_l \beta_l}a_{\beta_l}^{\dagger}$ and similar for the annihilation operators. Note that from this property it follows that the conjugation of an $n$-mode operator with these unitaries returns another $n$-mode operator. 

The orthogonal matrix $U^{(l)}$ that mixes the creation and annihilation operators is related to the rotation angles $\theta_{\alpha_l \beta_l}$ through the relation $U^{(l)} = e^{X^{(l)}}$, where $X^{(l)} = - {X^{(l)}}^T$ is an anti-symmetric matrix with off-diagonal elements $X^{(l)}_{\alpha_l \beta_l} = \theta_{\alpha_l \beta_l}$. $H_{1M}$ is thus already a  fast-forwardable fragment. The two-mode Hamiltonian $H_{2M}$ can similarly be decomposed into fast-forwardable fragments of the form,
\begin{align}
    H_{\nu} & = \mathcal{G}_{\nu} \left(\sum_{l > m} \sum_{i_l j_m} \lambda^{(\nu,lm)}_{i_l j_m} n_{i_l}n_{j_m}\right) {\mathcal{G}_{\nu}}^{\dagger} \notag \\
    & = \sum_{l > m}\sum_{\substack{\alpha_l \beta_l \\ \gamma_m, \delta_m}} g^{(\nu,lm)}_{\alpha_l \beta_l \gamma_m \delta_m} E^{\alpha_l}_{\beta_l} E^{\gamma_m}_{\delta_m}. \label{eq:cf_frag}
\end{align}
Thus, here, we use $\mathcal{G}_{\nu}$ as $\mathcal{U}_{\nu}$ in \eq{eq:Hnu}. 
The two-mode tensor corresponding to the $\nu$th fragment is defined as,
\begin{align}
    g^{(\nu,lm)}_{\alpha_l \beta_l \gamma_m \delta_m} = \sum_{i_l j_m} \lambda^{(\nu, lm)}_{i_lj_m} U^{(l)}_{i_l \alpha_l}U^{(l)}_{i_l \beta_l}U^{(m)}_{j_m \gamma_m}U^{(m)}_{j_m \delta_m}.
\end{align}
Note that the two-mode tensor of the Hamiltonian has four-fold symmetry, 
\begin{align}
    g^{(lm)}_{\alpha_l \beta_l \gamma_m \delta_m} &= g^{(lm)}_{\beta_l \alpha_l\gamma_m \delta_m} = g^{(lm)}_{\alpha_l \beta_l \delta_m \gamma_m}.
\end{align}
and so do the two-mode tensors of the fragments $g^{(\nu,l m)}_{\alpha_l \beta_l \gamma_m \delta_m}$. Fast-forwardable fragments of the form presented in Eq.~\eqref{eq:cf_frag} can be found by performing a non-linear optimization over the fragment parameters, $\{\lambda^{(\nu, l m)}_{i_l,j_m}, \theta^{(\nu,l)}_{\alpha_l \beta_l} \}$. In electronic structure theory, fast-forwardable two-electron fragments are found in an analogous fashion by using the Greedy Full Rank Optimization (GFRO) algorithm \cite{yen2021}. We adapt it here for the Christiansen form of the vibrational Hamiltonian. The steps of the algorithm are as follows:

\begin{enumerate}
    \item The one-mode part of the Hamiltonian, $H_{1M}$, is already a fast-forwardable fragment, label it $H_0$. Define $|g|_2\equiv ||g^{(l m)}_{\alpha_l \beta_l \gamma_m \delta_m}||_2$ as the $L_2$ norm of the two-mode tensor of the Hamiltonian. 

    \item \label{en:opt_cf} Find a fast-forwardable fragment $H_{\nu}$ of the form in Eq.~\eqref{eq:cf_frag} by optimizing the cost function $||g^{(l m)}_{\alpha_l \beta_l \gamma_m \delta_m} -g^{(\nu,l m)}_{\alpha_l \beta_l \gamma_m \delta_m}||_2$ over the fragment parameters $\{\lambda^{(\nu, l m)}_{i_l,j_m}, \theta^{(\nu,l)}_{\alpha_l \beta_l} \}$. The fragment parameters are restricted to be real.

    \item \label{en:store_cf} Store the fast-forwardable fragment $H_{\nu}$, and subtract it from the original Hamiltonian, $H \to H - H_{\nu}$. This has the effect of updating the two-mode tensors as, $g^{(l m)}_{\alpha_l \beta_l \gamma_m \delta_m} \to g^{(l m)}_{\alpha_l \beta_l \gamma_m \delta_m} - g^{(\nu,l m)}_{\alpha_l \beta_l \gamma_m \delta_m}$.

    \item \label{en: tol_cf} Repeat steps \ref{en:opt_cf} and \ref{en:store_cf} until the condition
    
    \begin{equation}
    \frac{||g^{(l m)}_{\alpha_l \beta_l \gamma_m \delta_m}||_2}{|g|_2} < \epsilon^{\text{CGF}}_{\text{frag}}
    \end{equation}
    is satisfied. For a sufficiently small tolerance, $\epsilon^{\text{CGF}}_{\text{frag}}$, the residual two-mode tensor $g^{(l m)}_{\alpha_l \beta_l \gamma_m \delta_m}$ can be discarded. Here $N_f$ is the number of two-mode fast-forwardable fragments found.

\end{enumerate}
From calculations on small systems we empirically determined that $\epsilon^{\text{CGF}}_{\text{frag}} = 0.05$ is sufficient to obtain an accuracy of $\sim 10$ cm$^{-1}$ in the eigenvalues. Note that $10\ \rm cm^{-1}$ corresponds to the typical resolution associated with the lineshape broadening width that is measured during an absorption spectrum. Appendix~\ref{app:cform_three_mode} presents the fragmentation scheme for the three-mode Hamiltonian. \\

\textbf{Hamiltonian mapped to Paulis:} \label{sec:cform_to_pauli}
An alternative route to direct fragmentation is to first map the Christiansen Hamiltonian to Pauli operators and then partition it into fragments. There are a number of different boson-to-Pauli mappings available \cite{sawaya2020}, but here we use the so-called direct or unary mapping. Refer to Appendix~\ref{app:unary_map} for details of the mapping. Note that within the Christiansen framework, physical vibrational wave functions lie in the single excitation per mode subspace of the full Hilbert space \cite{christiansen2004}: the number of excitations in modals corresponding to each mode will always add to one (Eq.~\eqref{eq:sepm_cond}). This allows us to truncate the occupation number for each operator to one while mapping to qubits, resulting in the mapping $E^{\alpha_l}_{\beta_l} = \frac{1}{4}\left(x_{\alpha_l}-iy_{\alpha_l}\right)\left(x_{\beta_l}+iy_{\beta_l}\right)$ for $ \alpha_l \neq \beta_l$, and $E^{\alpha_l}_{\alpha_l} = \frac{1}{2}\left(\mathbf{1}_{\alpha_l} - z_{\alpha_l}\right)$. The vibrational Hamiltonian thus gets mapped as,
\begin{align}
    H = \sum_{k=1}^{N_H} c_k P_k, \label{eq:pauli_ham}
\end{align}
where $P_k$ are $N_{Qb}=\sum_{l=1}^{M} N_l$ qubit Pauli operators, and $N_H$ is the number of Pauli terms in the Hamiltonian. We consider three popularly used fragmentation schemes for this Hamiltonian. The simplest considers each Pauli term as a fragment, and we refer to it as \gls{PF}. Note that for this fragmentation scheme, the Pauli index $k$ and the fragment index $\nu$ will be the same. The other two fragmentations schemes rely on gathering Pauli operators into groups of commuting Pauli operators \textemdash either \gls{FC} Pauli operators or the more restrictive \gls{QWC} Pauli operators. 
\begin{align}
    H & = \sum_{\nu} H_{\nu} = \sum_{\nu}\left(\sum_{k} c^{(\nu)}_k P^{(\nu)}_k\right) \notag \\
    & = \sum_{\nu}\mathcal{U}^{C}_{\nu}\left(\sum_{k} c^{(\nu)}_k Z^{(\nu)}_k\right){\mathcal{U}^{C}_{\nu}}^{\dagger} . \label{eq:pauli_fc_qwc}
\end{align}
Here, each fragment $H_{\nu}$ is a linear combination of either \gls{FC} or \gls{QWC} Pauli operators, $P^{(\nu)}_k$. Pauli products form an $\mathfrak{su}(2^{N_{Qb}})$ Lie algebra, its CSA consists of all $Z_k$ operators, with either a Pauli $Z$ operator or an identity on each qubit. Although all other commuting sets of operators can be obtained by Lie group conjugation
this is not feasible because of the exponential number of generators in $\mathfrak{su}(2^{N_{Qb}})$. Instead, the common practice is to use the 
Clifford group rotation $\mathcal{U}^{C}_{\nu}$ as $\mathcal{U}_{\nu}$ in \eq{eq:Hnu}: 
\gls{FC} uses the full Clifford group, while \gls{QWC} restricts it to only single-qubit subgroup. This is illustrated in the last equality in Eq.~\eqref{eq:pauli_fc_qwc}. The Clifford rotations conserve the number of Pauli products, and they can be obtained to transform any commuting Pauli product set to the CSA by techniques of symplectic linear algebra over the binary field
described in Ref.~\cite{yen_measuring_2020}. 
Since both \gls{FC} and \gls{QWC} schemes collect Pauli terms in the Hamiltonian into groups, the number of fragments in both cases is significantly smaller than $N_H$ in the \gls{PF} case. 

\subsection{Time Evolution} \label{sec:cf_te}

With the Hamiltonian fragmentations in hand, we now describe how to implement the time evolution propagator for each case. \\

\textbf{\glsentrylong{CGF}:} To implement the propagator of each fragment,
\begin{align}
     e^{-iH_{\nu}t} = \mathcal{G}_\nu e^{-iD_{\nu}t}  {\mathcal{G}_{\nu}}^{\dagger}, \label{eq:prop_cf}
\end{align}
using a quantum circuit, we need to map the excitation operators to qubits and decompose the modal rotation unitaries $\mathcal{G}_{\nu}$ and the diagonal fragment unitaries $e^{iD_{\nu}t}$ into gates. Here, we employ the unary boson-to-qubit mapping, with details in Appendix~\ref{app:unary_map}. The propagator for each fragment, and in turn the Trotterized propagator, can then be written as a product of Pauli rotations. Details are provided in Appendix~\ref{app:cf_mapping}. These Pauli rotations can be constructed using $R_z$ rotations and Clifford gates with the staircase algorithm \cite{Mansky2023}. The staircase algorithm can be understood as consisting of two steps. 

First, all operations involving $X$ or $Y$ Pauli operators for a given qubit can be transformed into $Z$ operations by conjugating with Hadamard and $S$ gates. This effectively expresses any Pauli exponential as a Clifford rotation conjugating a multi-qubit Pauli $Z$ rotation, e.g. $e^{i\theta x_1y_2z_3} = \mathcal U_C \ e^{i\theta z_1 z_2 z_3} \ \mathcal U_C^\dagger$. The second step then consists of decomposing the multi-qubit Pauli $Z$ rotation into a single-qubit $Z$ rotation that is conjugated by another Clifford operation. This Clifford operation can be constructed as a cascade of CNOTs connecting each of the qubits where a $Z$ operator appears with the qubit where the single $Z$ rotation will be done, e.g. $e^{i\theta z_1z_2z_3} = \mathcal U_C\  e^{i\theta z_3}\  \mathcal U_C^\dagger$. Overall this allows to decompose an arbitrary Pauli rotation as $e^{i\theta P} = \mathcal U_C \ R_{z}(2\theta)\ \mathcal U_C^\dagger$. 

Finally, we note that when performing a controlled implementation of the fragment propagator in Eq.~\eqref{eq:prop_cf}, the modal rotation unitaries $\mathcal{G}_\nu$ do not need to be controlled, only the individual diagonal fragment unitaries $e^{-iD_\nu t}$ do, which is a general property when controlling a conjugated unitary operator. This property can also be used for controlled time evolutions in different forms depending on the associated implementation of each $e^{-iD_\nu t}$ operator. \\

\textbf{Hamiltonian mapped to Paulis:} Alternatively, if the Christiansen Hamiltonian is first mapped to Pauli operators \eq{eq:pauli_ham}, the propagator can be constructed in one of three ways.  In the \gls{PF} scheme, where each Pauli is taken to be an individual fragment, the propagator of each fragment $e^{ic_{\nu} P_{\nu}t}$ is a Pauli rotation and can be constructed using $R_z$ rotations and Clifford gates with the staircase algorithm \cite{Mansky2023} as mentioned before. Controlling the Pauli rotation can simply be done by controlling the $R_z$ rotation within the staircase algorithm, since the Clifford operations do not need to be controlled because they occur in conjugate pairs. Lastly, for the \gls{FC} or \gls{QWC} schemes, the last equality in Eq.~\eqref{eq:pauli_fc_qwc} illustrates that the propagator for each fragment $e^{iH_{\nu}t}$ can be implemented by conjugating $R_z$ rotations with Clifford gates. 

\subsection{Trotter Error Analysis}
In the Christiansen form, the Trotter error is estimated using Eq.~\eqref{eq:trial_state_trot_error}, where the contribution from the $n$th eigenstate is given in Eq.~\eqref{eq:nth_trot_error}. The operator $\Theta_2$ differs for each fragmentation scheme used. For the \gls{PF} scheme, each fragment, $H_{\nu}$ in Eq.~\eqref{eq:V2}, is a Pauli word in the Hamiltonian. For the \gls{FC} or \gls{QWC} schemes, each fragment is a linear combination of \gls{FC} or \gls{QWC} Pauli operators, as shown in Eq.~\eqref{eq:pauli_fc_qwc}. For the \gls{CGF} scheme, each fragment is as defined in Eq.~\eqref{eq:cf_frag} mapped to Pauli operators. As will be discussed in Sec.~\ref{sec:results}, we estimate T gate costs for a selection of several molecules and also demonstrate our approach by calculating an infrared-spectrum for the H$_2$S molecule. For these calculations, we use the states  $\mu\ket{{\Phi}_0}$ or $\mu\ket{\tilde{\Phi}_0}$ to estimate the Trotter error, where $\ket{{\Phi}_0}$ and  $\ket{\tilde{\Phi}_0}$ are the \gls{HO} and \gls{VSCF} basis ground state respectively, and $\mu$ is the dipole operator written in the respective \gls{HO}/\gls{VSCF} basis.

\subsection{Resource Estimates}
\label{sec:cform_resource}
Here we calculate the number of T gates required to implement one second-order Trotter oracle $e^{-iH\Delta t }$, and a long time evolution for time $t_{\text{max}}$ consisting of $L_{\text{max}}$ Trotter steps, using various fragmentation schemes in the Christiansen form. Note that the cost of implementing $L_{\text{max}}$ Trotter steps is not just $L_{\text{max}}$ times the cost of each Trotter oracle due to cost reductions that can be obtained from merging unitaries appearing at the end of one Trotter oracle and the beginning of the next. Each Trotter oracle can be implemented using a product of Pauli rotations and Clifford gates. Pauli rotations can be constructed using $R_z$ rotations and Clifford gates with the staircase algorithm \cite{Mansky2023}. We use the method of Ref.~\cite{Bocharev2015} to estimate the number of T gates per $R_z$ gate using a synthesis accuracy of $10^{-10}$, such that each $R_z$ requires approximately $50$ T gates. As mentioned before, if the algorithm demands a controlled time evolution, the only modification will be needing to control the relevant $R_z$ gates. The resource counts for that case are modified in a straightforward way: each controlled $R_z$ gate can then be implemented using two uncontrolled $R_z$ rotations plus two CNOTs \cite{optimal_t_decomps}. Combining this with the above cost of a single uncontrolled $R_z$ gate, the cost of each controlled $R_z$ then is around $100$ T gates. In both cases, to estimate the final T gate cost, we need to calculate the $R_z$ gate count. Note that for our calculations we chose the number of modals for each mode to be the same, namely $N_l = N$ for all $l \in [1,M]$, although the general case follows a similar reasoning.

\textbf{\glsentrylong{CGF}:} For the \gls{CGF} scheme with $2$-mode coupling, the number of $R_z$ gates per Trotter oracle is 
\begin{multline}
N_{R_z}^{CGF} = 2MN + 2MN(N-1)(N_f+1)  \\
+ MN\left( 1 + \frac{1}{2}N(M-1)\right)\left(2N_f -1 \right),
\end{multline}
where $N_f$ is the number of $2$-mode fragments. This includes contributions from the unitaries that diagonalize each fragment, and from the diagonal $1-$ and $2-$mode fragments themselves. Appendix~\ref{app:cf_res_est} presents more details on how this cost is calculated, including the cost of long-time evolution with $L_{\text{max}}$ Trotter steps and generalizations to higher mode couplings. Asymptotically, the cost per Trotter oracle for the \gls{CGF} scheme scales as,
\begin{align}
    \mathcal{C}_{\text{CGF}} \sim \mathcal{O}\left(N_f M^n N^n\right),
\end{align}
where $n$ is the number of mode couplings included in the Hamiltonian and $N_f$ is the number of $n$-mode fragments. In close analogy to the compressed double factorization method for the electronic structure Hamiltonian \cite{rocca2024}, the number of fragments $N_f$ is expected to scale as at most $\mathcal{O}\left(MN^2\right)$ for the $n=2$ case, but will very likely be much smaller in practice.

\textbf{Hamiltonian mapped to Paulis:} In the \gls{PF} method, the number of $R_z$ gates per second-order Trotter oracle will be,
\begin{align}
    N_{R_z}^{PF} = 2N_H - 1,     
\end{align}
where $N_H \sim \mathcal{O}\left( M^n N^{2n}\right)$ \cite{trenev2025}, and $n$ is the order of mode-couplings included in the Hamiltonian. Similarly, even in the \gls{FC} and \gls{QWC} fragmentation schemes, although the Pauli terms are first grouped and then exponentiated in groups, the total number of Pauli rotations will still be very similar to the PF case, with small reductions coming from the last fragment being implemented only once in the second-order Trotter product formula. Details are provided in Appendix~\ref{app:pauli_est}, along with costs for long time evolution. Asymptotically, the cost per Trotter oracle for all three schemes scales as,
\begin{align}
    \mathcal{C}_{\text{PF/QWC/FC}} \sim \mathcal{O}\left( M^n N^{2n}\right) .
\end{align}

\section{Real space form} \label{sec:real-space}
\subsection{Hamiltonian}
Once the $n$-mode expansion for the vibrational \gls{PES} has been obtained (Eq.~\eqref{eq:n_mode}), we can perform a Taylor series expansion of the $\mathcal{V}^{(n)}$ terms. Since we are expanding around an equilibrium position the first-order derivatives are trivially $0$, as shown in Eq.~\eqref{eq:eq_jacobian}, meaning there will be no terms linear in the $q_i$'s. We also need include the kinetic energy operator, which can be generally expressed as $T_{\rm vib} = \sum_{ij} K_{ij} p_ip_j$ where the $K_{ij}$ matrix is diagonal for canonical normal modes (Eq.~\eqref{eq:kin_normal}) or has a more general form for localized modes (Eq.~\eqref{eq:final_kinetic}). Overall, this yields the real space Hamiltonian form
\begin{align}
    H &= T+V \\
    &= \sum_{ij} K_{ij} p_ip_j + \sum_{ij}V^{(2)}_{ij} q_iq_j \nonumber \\
    &\ \ + \sum_{ijk} V^{(3)}_{ijk} q_iq_jq_k + \sum_{ijkl} V^{(4)}_{ijkl} q_iq_jq_kq_l. \label{eq:taylor_ham}
\end{align}
Here we have truncated the Taylor expansion of the \gls{PES} at fourth order as commonly done in the literature under the name ``quartic force field'' \cite{qff_1,qff_2,qff_3}. In general, higher order expansions can be used for increased accuracy. Note that in practice these expansion coefficients can also be obtained by calculating high-order derivatives of the \gls{PES} around the equilibrium geometry, bypassing the need to perform the $n$-mode expansion.

\begin{figure}
    \centering
    \includegraphics[width=1\linewidth]{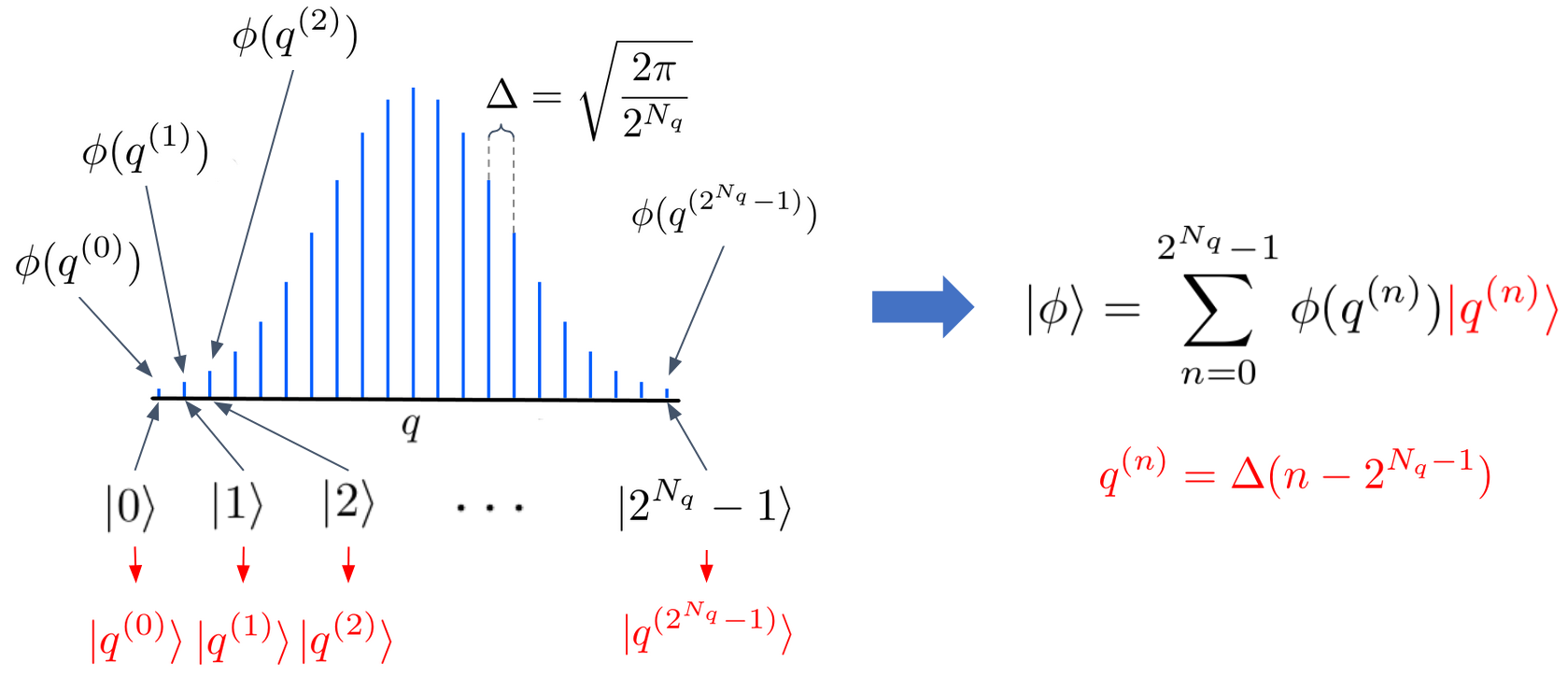}
    \caption{Real space form \cite{macridin2018a,macridin2018b} of single vibrational mode $q$. Each of the $2^{N_q}$ integers associated with the $N_q$-qubit computational basis states are associated with a different point in the grid discretizing $q$. The wavefunction $\ket{\phi}$ will then be represented by a coherent superposition over computational basis states with associated coefficients $\phi(q^{(n)})$, namely $\ket{\phi} = \sum_n \phi(q^{(n)}) \ket{q^{(n)}}$. Adapted with permission from Ref.~\cite{loaiza2025}.}
    \label{fig:rs_grid}
\end{figure}

We now discuss how the vibrational space is encoded in the real space form, as originally introduced in Refs.\cite{macridin2018a,macridin2018b} and used for vibrational and vibronic simulation in Refs.~\cite{loaiza2025,motlagh_vibronic}. The real space form of the Hamiltonian uses a qubit register with $N_q$ qubits for each vibrational mode. Each of the vibrational coordinates is discretized using a uniform grid with $2^{N_q}$ points. Thus, a given computational basis state with associated integer value $n$ encodes the value of the vibrational wavefunction at the associated grid point, which we can write as the action of the vibrational mode operator
\begin{equation} \label{eq:rs_grid}
    \hat q_l \ket{n} = q_l^{(n)}\ket{n}=(n-2^{N_q-1})\Delta \ket{n},
\end{equation}
where $\Delta=\sqrt{\pi\cdot2^{1-N_q}}$ corresponds to the spacing between grid points with associated total width $\sqrt{\pi\cdot2^{N_q+1}}$. A visual representation of how the grid is constructed is shown in Fig.~\ref{fig:rs_grid}.
Note that as discussed in Refs.\cite{macridin2018a,macridin2018b}, through the use of the Nyquist-Shannon theorem it is guaranteed that a logarithmic number of qubits $N_q$ is enough to accurately represent the vibrational space, having that in practice a choice of $N_q=4$ or $5$ for each vibrational mode is already enough to obtain high-accuracy eigenvalues from the associated Hamiltonian.

\subsection{Fragmentation} \label{sec:rs_frag}
The Heisenberg algebra $\mathfrak{h}(M)$ is generated by the operators $\{p_j, q_j, \bm{1}\}_{j=1}^M$. The real space Hamiltonian in Eq.~\eqref{eq:taylor_ham}, which is a polynomial of these operators, thus belongs to the universal enveloping algebra of the Heisenberg algebra. 
The Heisenberg algebra is nilpotent and this does 
not allow us to simply use its CSA to find fragments \textemdash the CSA for a nilpotent algebra 
is the entire algebra. Instead, the two MASAs representable by normal operators 
are employed to formulate diagonalizable fragments: $\{1,q_j\}_{j=1}^{M}$ 
and $\{1,p_j\}_{j=1}^{M}$. Another problem with nilpotent Lie algebras is that  corresponding Heisenberg Lie group elements cannot map elements of $\{p_j, \bm{1}\}_{j=1}^M$ to those of $\{q_j,\bm{1}\}_{j=1}^{M}$ 
by conjugation. For connecting these subalgebras of $\mathfrak{h}(M)$
we use the \gls{QFT} unitary as $\mathcal{U}_{\nu} = \mathtt{Q_M}$ in \eq{eq:Hnu}, whose action is defined as
\bea
\mathtt{Q_M}^\dagger q_j \mathtt{Q_M} &=& p_j \\
\mathtt{Q_M}^\dagger p_j \mathtt{Q_M} &=& -q_j.
\eea
This approach allows us to identify that the potential term $V(\bm{q})$, and the kinetic term $T(\bm{p})$ are already fast-forwardable fragments of the real space Hamiltonian. Furthermore, since the real space approach essentially uses the position basis, the potential fragment is already in diagonal form. Thus the real space Hamiltonian can be decomposed into fast-forwardable fragments as,
\begin{align}
    H = \mathtt{Q_M}^\dagger \left(\sum_{ij} K_{ij} q_iq_j\right) \mathtt{Q_M} + V(\bm{q}). \label{eq:rs_frags}
\end{align}
Details of the implementation of the propagator in the real space approach are presented in next section.

\subsection{Time Evolution}
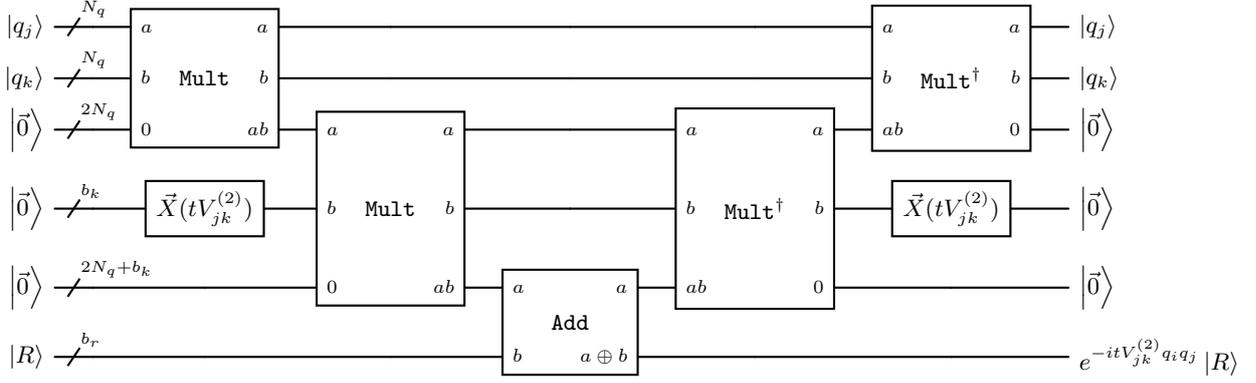
\begin{figure*}
    \centering
    \begin{quantikz}[row sep=0.4cm]
		 \lstick{$\ket{q_j}$} & \qwbundle{N_q} & \gate[3]{\hspace{0.5cm}\mathtt{Mult}\hspace{0.5cm}} \gateinput{$a$} \gateoutput{$a$} &&&& \gate[3]{\hspace{0.5cm}\mathtt{Mult}^\dagger\hspace{0.5cm}} \gateinput{$a$} \gateoutput{$a$} & \rstick{$\ket{q_j}$} \\ 
		\lstick{$\ket{q_k}$} & \qwbundle{N_q} & \gateinput{$b$} \gateoutput{$b$} &&&& \gateinput{$b$} \gateoutput{$b$} & \rstick{$\ket{q_k}$} \\
		\lstick{$\ket{\vec 0}$} & \qwbundle{2N_q} & \gateinput{$0$} \gateoutput{$ab$} & \gate[3]{\hspace{0.5cm}\mathtt{Mult}\hspace{0.5cm}}\gateinput{$a$} \gateoutput{$a$} &  & \gate[3]{\hspace{0.5cm}\mathtt{Mult}^\dagger\hspace{0.5cm}}\gateinput{$a$} \gateoutput{$a$} & \gateinput{$ab$} \gateoutput{$0$} & \rstick{$\ket{\vec 0}$} \\
		\lstick{$\ket{\vec 0}$} & \qwbundle{b_k} & \gate{\vec X(tV^{(2)}_{jk})} & \gateinput{$b$} \gateoutput{$b$} && \gateinput{$b$} \gateoutput{$b$} & \gate{\vec X(tV^{(2)}_{jk})} & \rstick{$\ket{\vec 0}$} \\
		\lstick{$\ket{\vec 0}$} & \qwbundle{2N_q+b_k} && \gateinput{$0$} \gateoutput{$ab$} & \gate[2]{\hspace{0.5cm}\mathtt{Add}\hspace{0.5cm}} \gateinput{$a$} \gateoutput{$a$} & \gateinput{$ab$} \gateoutput{$0$} && \rstick{$\ket{\vec 0}$} \\
		\lstick{$\ket{R}$} & \qwbundle{b_r} &&& \gateinput{$b$} \gateoutput{$a\oplus b$} &&& \rstick{$e^{-itV_{jk}^{(2)} q_i q_j}\ket{R}$}
	\end{quantikz}
    \caption{Quantum arithmetic-based circuit for implementing the exponential operator of a monomial of $q_j$'s, namely $e^{-itV_{jk}^{(2)}q_jq_k}$. For compactness we denote $\ket{q_j}$ as the qubit register encoding the vibrational coordinate $j$. The $\mathtt{Mult}$ directives correspond to out-of-place multipliers \cite{wang2025comprehensive}, while the $\mathtt{Add}$ circuits to out-of-place (or Gidney) adders for the phase gradient technique \cite{gidney2018}. The $\vec X(V_{jk}^{(2)})$ gates correspond to a series of Pauli X gates which load the binary representation of the $tV_{jk}^{(2)}$ coefficient in the associated register with $b_k$ qubits. Operations after the modular addition effectively unload the multiplication results from the used additional qubit registers.}
    \label{fig:rs_exponential}
\end{figure*}

For implementing the time evolution, we start by showing how to implement the potential term, with the understanding that the kinetic term will use the same machinery. Note that the potential term is a function of $q_i$'s, making it already diagonal in the real space form: the computational basis functions are eigenfunctions of the $q_i$ operators [Eq.~\eqref{eq:rs_grid}]. The associated exponential of this fragment can then be written as
\begin{align}
    e^{-iVt}  
    &= \left(\prod_{jk} e^{-it V^{(2)}_{jk}q_jq_k}\right)\left(\prod_{jkl} e^{-it V^{(3)}_{jkl}q_jq_kq_l}\right) \notag \\
    & \times \left(\prod_{jklm} e^{-it V^{(4)}_{jklm}q_jq_kq_lq_m}\right). \label{eq:rs_monomials}
\end{align}
The implementation of this operator thus requires the exponentiation of monomials of $q_i$ operators. Each of these monomials is efficiently implemented through the use of quantum arithmetic and the phase gradient technique \cite{gidney2018}. The basic idea of the phase gradient technique is to use a resource state
\begin{equation}
    \ket{R} = \frac{1}{\sqrt{2^{b_r}}} \sum_{k=0}^{2^{b_r}-1} e^{ik\phi_{b_r}} \ket{k},
\end{equation}
that has been defined over a register with $b_r$ qubits with $\phi_{b_r}=2\pi\ell/2^{b_r}$. We have here defined the integer $\ell$ which incorporates the information of the scale associated with the integer $k$. It can be shown that the modular addition of this state with another register yields
\begin{equation}
    \mathtt{Add}\{\ket{n}\ket{R}\} = e^{-in\phi_{b_r}}\ket{n}\ket{R},
\end{equation}
where $n$ simply corresponds to an integer that is encoded in an associated computational basis state. The constant $\phi_{b_r}$ thus connects the value of the integer number $n$ with an associated phase value $n\phi_{b_r}$. In practice this means that we will need a different resource state for each different monomial degree, which in this case would require three different resource registers associated to the second-, third-, and fourth-degree monomials appearing in Eq.~\eqref{eq:rs_monomials}. For a monomial of degree $d$, the associated constant $\phi_{b_r}$ will then be associated to a $b_r$-bit approximation of $\Delta^d$, which as shown in Eq.~\eqref{eq:rs_grid} is necessary for connecting the value of the integer encoded by the qubits to the associated application of the $q_j$ position operators. 

The basic idea for implementing the time evolution associated with one of the monomials, e.g. $e^{-itV_{jk}^{(2)} q_jq_k}$, is to perform a bit encoding of the coefficient $tV_{jk}^{(2)}q_jq_k$ in a quantum register, which then implements the desired exponential through a modular addition with a resource state. For monomials of higher degree more intermediate registers are used to store intermediate multiplication results: note that two multiplication operations are required for each additional degree, corresponding to one computing and one uncomputing operation. Note that in practice we can also include the scale of the time step $t$, alongside a scale for the monomial coefficients of that given degree in the $\phi_{b_r}$ constant of the resource state, such that the $b_k$ qubits used in the coefficient register encoding e.g. $V_{jk}^{(2)}$ contain as much information as possible of this coefficient with respect to all other second-degree coefficients. A circuit showing the implementation of this term is shown in Fig.~\ref{fig:rs_exponential}. \\

Having discussed how to implement a single monomial term, and how the time evolution by the potential energy can be expressed as a product of such operations, it now becomes clear how the implementation of the full potential energy evolution $e^{-iVt}$ can be implemented by this procedure. There are now two more details to be covered for a complete specification of the real-space-based time evolution implementation. First, we will discuss the coefficient caching technique \cite{loaiza2025,motlagh_vibronic}. The basic idea is to notice that the intermediate coefficients that are calculated through the quantum multiplications often appear in different monomials, meaning that different monomials might be computed from some term in common before the uncomputation step. To make this point clearer, we consider the three monomials $V^{(2)}_{jk}q_jq_k$, $V^{(3)}_{jjk}q_j^2q_k$, and $V^{(4)}_{jklm}q_jq_kq_lq_m$. For simplicity we refer to a state $\ket{q_j}$ as a qubit register encoding the vibrational coordinate $j$. All three of the mentioned monomials will require the intermediate storage of an integer associated with the $\ket{q_jq_k}$ state. As such, the multiplication of the $\ket{q_j}$ and $\ket{q_k}$ registers does not need to be implemented a total of six times (three computations and three uncomputations), but instead only two applications are enough. In practice, finding the optimal caching strategy can be modelled as a tree traversal problem, for which many efficient heuristics exist \cite{tree_traversal_1,tree_traversal_2}. We also note that it is possible to increase the number of intermediate registers, effectively increasing the required number of additional qubits in exchange for being able to store more cached coefficients throughout the calculation, lowering the overall number of multiplications. \\

The last consideration for fully specifying the time evolution operator in this form corresponds to the implementation of the exponential of the kinetic energy operator. The multimode \glsentrylong{FT} unitary $\mathtt{Q_M}$ corresponds to the diagonalizing unitary for the kinetic energy operator, as shown in Eq.~\eqref{eq:rs_frags}. It is composed of shifted \glspl{QFT} over individual modes
\begin{equation}
    \mathtt{Q_M} = \prod_{j=1}^M \mathtt{QFT}_{j,s},
\end{equation}
where $\mathtt{QFT}_{j,s}$ stands for a shifted \gls{QFT} over the qubit register encoding the vibrational mode $j$. This entails the application of the regular \gls{QFT} followed by a bit flip of the most significant qubit in the register. This is completely analogous to the often applied shift for a discrete \gls{FT} in classical computing, effectively centering the \gls{FT} in some range $[-a,a)$ instead of $[0,2a)$. It acts as
\begin{align} \label{eq:q_to_p}
     \mathtt{QFT}_{j,s}^\dagger \cdot q_j \cdot \mathtt{QFT}_{j,s} &= p_j \\
     \mathtt{QFT}_{j,s}^\dagger \cdot p_j \cdot \mathtt{QFT}_{j,s} &=-q_j.
\end{align}
After this rotation using the $\mathtt{Q_M}$ unitary, we simply require the implementation of a second-degree polynomial of $q_j$'s, which can be applied using the same technique that was just described for the potential energy terms. Having shown how to implement both the time evolutions for the potential and kinetic energy operators, the specification of the time evolution operator for the real space form is now complete. \\

Finally, we note that for cases where the controlled time evolution is required, this can be done by simply adding a control to the addition operation implementing the phase gradient, which will incur an additional cost of $4$ T gates per-addition.

\subsection{Trotter Error Analysis}
As discussed in the following section, the real space and bosonic forms of the Hamiltonian both come from the power series expansion of the vibrational Hamiltonian, as shown in Eq.~\eqref{eq:taylor_ham}. In addition, all algebraic properties of the bosonic form will be faithfully captured by the real space form as long as the operations do not hit the limits of the grid: all properties of the real space and bosonic forms are identical in a low energy subspace that is determined by the number of qubits used for discretizing the vibrational coordinates. In practice this number is chosen as to accurately capture all properties of interest, guaranteeing that the vibrational wavefunction never hits the limits of the discretized grid. The Trotter error analysis of both the real space and bosonic approaches is thus identical for the chosen basis sizes. Due to computational efficiency, the Trotter error analysis of the real space approach is done using the bosonic approach, which is described alongside the bosonic form in the next section.

\subsection{Resource Estimates}
We now present the number of T gates required to implement a time evolution associated to $e^{-i\hat H t_{\rm max}}$ using $L_{\rm max}$ second-order Trotter steps, each for a time of $\Delta t=t_{\rm max}/L_{\rm max}$. We first start by finding the complexity of a single second-order Trotter oracle. Full details on how this complexity is obtained are shown in Appendix~\ref{app:rs_cost}. The basic idea is to notice that the cost will be governed by the implementation of each coefficient in the Taylor expansion of the Hamiltonian, with the cost of each of these coefficients requiring a series of quantum arithmetic operations: in particular, implementing a term of $\ell$-th degree requires $\ell$ multiplications and one addition. The cost of implementing each $\ell$-th degree term can be thus shown to be given by
\begin{align}
    \mathcal{C}^{(\ell)} &= 2N_q(N_q\ell(\ell-1)+\ell(2b_k-1)+3) +4b_k-4 \\
    &\sim\mathcal{O}(\ell^2N_q^2) \nonumber
\end{align}
T gates. In the worst-case scenario where all possible coefficients are non-zero, we have that the total number of $\ell$-th degree coefficients is given by
\begin{equation}
    \mathcal N^{(\ell)} = \begin{cases}
        0,\ \textrm{if}\  \ell < 2 \\
        \sum_{j=1}^\ell \binom{\ell-1}{\ell-j}\binom{M}{j},\ \textrm{if}\  2\leq\ell\leq n \\
        \sum_{j=1}^n \binom{n-1}{n-j} \binom{M}{j},\ \textrm{if}\ n<\ell\leq d,
    \end{cases}
\end{equation}
where  $n$ corresponds to the degree of the $n$-mode expansion and we have assumed $n\leq d$.  From this we have that the full implementation of this approach requires at most
\begin{equation}
    \mathcal{C}_{\rm real\ space} = \sum_{\ell=2}^d \mathcal{C}^{(\ell)} \mathcal{N}^{(\ell)}
\end{equation}
T gates. Noting that the complexity will be dominated by the $\ell=d$ terms and that $\mathcal{N}^{(d)}\sim \mathcal{O}(M^n)$, we recover the scaling of this approach as
\begin{equation}
    \mathcal{C}_{\rm real\ space} \sim \mathcal{O}(d^2 N_q^2 M^n),
\end{equation}
while requiring 
\begin{equation}
    \mathcal{Q}_{\rm real\ space} = \frac{N_q(d^2+3d-2)}{2}+2b_k+(d-1)b_r
\end{equation}
additional ancilla qubits. The $(d-1)b_r$ contribution comes from the resource states for each $\ell=2,\cdots,d$ possible monomial powers, while the other ancillas correspond to the intermediary registers where the quantum arithmetic results are stored. Note that this analysis does not consider the cost improvements from using the coefficient caching technique, having that both the complexity and number of ancilla qubits will remain the same: details on the caching technique may be found in Ref. \cite{loaiza2025}. \\

Finally, considering all $L_{\rm max}$ Trotter steps using a second-order Trotter formula will require $L_{\rm max}$ applications of the potential energy evolution oracle and $L_{\rm max}+1$ applications of the kinetic evolution oracle. Here we have used the fact that there is a telescopic cancellation when considering the product of $L_{\rm max}$ second-order Trotter oracles. Namely, for a single second-order Trotter oracle $U_{\Delta t} =e^{-iT \Delta t/2}e^{-iV\Delta t}e^{-iT \Delta t/2}$, consecutive applications will combine the kinetic energy evolutions, yielding $(U_{\Delta t})^{L_{\rm max}} = e^{-iT \Delta t/2} \left(\prod_{j=1}^{L_{\rm max}-1} e^{-iV\Delta t} e^{-iT\Delta t}\right) e^{-iV\Delta t} e^{-iT \Delta t/2} $.

\section{Bosonic form } \label{sec:bosonic}
\subsection{Hamiltonian}

The vibrational Hamiltonian used in the bosonic form is obtained from the real space Hamiltonian in Eq.~\eqref{eq:taylor_ham}. For this purpose, the coordinate ($q_j$) and momentum ($p_j$) operators are expressed in terms of bosonic creation and annihilation operators,
\begin{align}
    q_j &= \frac{1}{\sqrt{2}}(b_j^{\dagger} + b_j),  & p_j &= \frac{i}{\sqrt{2}}(b_j^{\dagger} -  b_j).
\end{align}
These operators follow the canonical commutation relations $[b_i,{b}_j^{\dagger}] = \delta_{ij}$, $[b_i,{b}_j] = 0$, and $[b_i^{\dagger},{b}_j^{\dagger}]$ = 0. The vibrational Hamiltonian expressed in terms of  bosonic creation and annihilation operators is given by
\begin{align}
    {H} & = \sum_{ij=1}^M\frac{-K_{ij}}{2}\left({b}_i^{\dagger}-{b}_i\right)\left({b}_j^{\dagger}-{b}_j\right) \notag \\
    & + \sum_{ij=1}^M\frac{V^{(2)}_{ij}}{2}\left({b}_i^{\dagger}+{b}_i\right)\left({b}_j^{\dagger}+{b}_j\right) \notag \\
    & + \sum_{ijk=1}^M \frac{V^{(3)}_{ijk}}{2\sqrt{2}}\left({b}_i^{\dagger}+{b}_i\right)\left({b}_j^{\dagger}+{b}_j\right)\left({b}_k^{\dagger}+{b}_k\right) \notag  \\
    & + \sum_{ijkl=1}^M \frac{V^{(4)}_{ijkl}}{4}\left({b}_i^{\dagger}+{b}_i\right)\left({b}_j^{\dagger}+{b}_j\right)\left({b}_k^{\dagger}+{b}_k\right)\left({b}_l^{\dagger}+{b}_l\right).
\end{align}
Similar to the real space approach, the Hamiltonian can also be obtained by calculating higher-order derivatives of the \gls{PES} around the equilibrium geometry, bypassing the need to perform the $n$-mode expansion. Note that we have retained terms up to fourth order in the PES, but higher-order terms can also be considered. As a result, the Hamiltonian contains terms that are up to quartic in the ladder operators.

\subsection{Fragmentation} \label{sec:bosonic_frag}
Bosonic Hamiltonians with terms up to quadratic order can be diagonalized  using a Bogoliubov transformation as \cite{Bogoliubov2009},
\begin{align}
     H_{\text{quad}} & = \sum_{lm} A_{lm} {b}_l^{\dagger}{b}_m + \frac{1}{2}B_{lm}{b}_l^{\dagger}{b}_m^{\dagger} + \frac{1}{2}B^{*}_{lm}{b}_l{b}_m \notag \\
     & + \sum_{l} C_l b_l^{\dagger} + C_l^* b_l \notag  \\
     & = \mathcal{B} \left(\sum_{l} \epsilon_l n_l + c \bm{1} \right) \mathcal{B}^{\dagger}. \label{eq:diag_Hlq}
\end{align}
Here, matrices $A$ and $B$ satisfy $A = A^{\dagger}$ and $B = B^{T}$ respectively, to ensure that the Hamiltonian is Hermitian; $c$ is a constant, $\bm{1}$ is the identity operator, $n_j = {b}_j^{\dagger}{b}_j $ is the number operator for the $j$th mode.
The Bogoliubov unitary $\mathcal{B} = \mathtt{D} \, \mathtt{G}$, is a product of displacement and quadratic Gaussian unitaries,
\begin{align}
    \mathtt{D} & = e^{\sum_{l}\gamma_{l} b_l - \gamma_l^* b^{\dagger}_l}, \label{eq:u_disp} \\
    \mathtt{G} & = e^{\sum_{lm} \alpha_{lm} {b}_l^{\dagger}{b}_m + \frac{1}{2}\left(\beta_{lm}{b}_l^{\dagger}{b}_m^{\dagger} + \beta_{lm}^{*}{b}_l{b}_m\right)}
\end{align}
that displace, and mix and squeeze bosonic modes respectively. $\bm{\alpha}^{\dagger} = -\bm{\alpha}$ and $\bm{\beta}^T = \bm{\beta}$ guarantee that the generator of the quadratic Gaussian unitary is anti-Hermitian. The parameters for the diagonalizing Bogoliubov unitary $\{\gamma_l, \alpha_{lm}, \beta_{lm}\}$ can be obtained from the Hamiltonian parameters $\{A_{lm}, B_{lm}, C_l \}$ \cite{Bogoliubov2009, malpathak2025}. 

The terms in the parenthesis of Eq.~\eqref{eq:diag_Hlq} are the diagonal quadratic Hamiltonian because the set $\{n_j,\bm{1}\}$  is the CSA of the inhomogeneous symplectic Lie algebra $\mathfrak{isp}(2N,\mathbb{C})$: $\{{b}_l^{\dagger}{b}_m, {b}_l^{\dagger}{b}_m^{\dagger}, {b}_l{b}_m, b_l^{\dagger},b_l,\bm{1}\}$. The Bogoliubov transformation, $\mathcal{U}_\nu = \mathcal{B}_{\nu}$ in \eq{eq:Hnu}, can be used here to produce fast-forwardable fragments by conjugating polynomials of the CSA, 
\begin{align}
    H_{\nu} & = \mathcal{B}_{\nu} \left(\sum_{lm} \eta^{(\nu)}_{lm} {n}_l {n}_m \right) {\mathcal{B}_{\nu}}^{\dagger} \label{eq:sol-bos-frags},
\end{align}
where $\eta_{lm}^{(\nu)}$ are fragment parameters and $\mathcal{B}_{\nu}$ denotes the Bogoliubov unitary corresponding to the $\nu$th fragment. These Bogoliubov unitaries are parametrized by $\{\alpha_{lm}^{(\nu)},\beta_{lm}^{(\nu)}, \gamma_{l}^{(\nu)} \}$. Details are provided in Appendix~\ref{app:bf_mapping}. The vibrational Hamiltonian is then decomposed into fast-forwardable fragments of the form in Eq.~\eqref{eq:sol-bos-frags} by obtaining the fragments one at a time using a greedy algorithm which is analogous to the GFRO algorithm used for the \gls{CGF} scheme presented in Sec.~\ref{sec:cform_frag}. This fragmentation scheme was recently used by some of us in Ref.~\cite{malpathak2025} to develop algorithms for vibrational spectroscopy and dynamics on bosonic quantum devices, where it is presented in detail. The steps of the algorithm are, 

\begin{enumerate}

    \item Collect the coefficients of the cubic and quartic terms in the Hamiltonian, henceforth referred to as $c_j$.

    \item \label{en:opt_bf} Find a fast-forwardable fragment $H_{\nu}$ of the form in Eq.~\eqref{eq:sol-bos-frags} by optimizing the cost function $\sum_j (c_j -c_j^{(\nu)})^2$ over the fragment parameters $\{\eta_{lm}^{(\nu)},\alpha_{lm}^{(\nu)},\beta_{lm}^{(\nu)},\gamma_l^{(\nu)}\}$, where $c_j^{(\nu)}$ are the coefficients of the cubic and quartic terms of the fast-forwardable fragment. The fragment parameters are restricted to be real.

    \item \label{en:store_bf} Store the fast-forwardable fragment $H_{\nu}$, and subtract it from the original Hamiltonian, $H \to H - H_{\nu}$. This has the effect of changing the cubic and quartic coefficients in the Hamiltonian as $c_j\to c_j - c_j^{\nu}$, while also modifying the linear, quadratic and constant terms.

    \item \label{en: tol_bf} Repeat steps \ref{en:opt_bf} and \ref{en:store_bf} until the residual of the coefficients of the cubic and quartic terms $\sqrt{\sum_{j}  (c_j -\sum_{k=1}^{N_f} c_j^{(k)})^2} < \epsilon^{\text{BF}}_{\text{frag}}$ . For a sufficiently small tolerance, the remaining cubic and quartic terms can be discarded. $N_f$ is the number of quartic fast-forwardable fragments found.

    \item \label{en: quad} The remaining Hamiltonian, $H - \sum_{\nu=1}^{N_f} H_{\nu}$, only contains up to quadratic terms, and can be diagonalized using a Bogoliubov transform. This fragment is referred to as $H_0$. 
\end{enumerate}
As a result, the Hamiltonian is decomposed into fast-forwardable fragments as $H = \sum_{\nu = 0}^{N_f} H_{\nu}$, with $H_{\nu = 0}$ being a quadratic fragment of the form in Eq.~\eqref{eq:diag_Hlq}, and the rest being quartic fragments of the form in Eq.~\eqref{eq:sol-bos-frags}.  From calculations on small systems we empirically determined that $\epsilon^{\text{BF}}_{\text{frag}} = 1$ cm$^{-1}$ is sufficient to obtain an accuracy of $\sim 10$ cm$^{-1}$ in the eigenvalues. From now on, we refer to these fragmentation scheme as \gls{BF}. Lastly, we note that the vibrational Hamiltonian in the bosonic form can also be fragmented after being mapped to Pauli operators as demonstrated in Sec.~\ref{sec:cform_frag}. However, similar to the Christiansen case, we expect these to be more expensive than the greedy fragmentation method. Since the bosonic greedy fragmentation scheme is already more expensive than the other schemes considered in this work, fragments of the bosonic Hamiltonian mapped to Paulis are expected to be even more expensive and are thus not considered in this work.

\subsection{Time Evolution}
To implement the propagator of each fragment, 
\begin{align}
    e^{-iH_{\nu}t} = \mathcal{B}_{\nu} e^{-iD_{\nu}t}  {\mathcal{B}_{\nu}}^{\dagger}, \label{eq:prop_bf}
\end{align}
as a quantum circuit, the Bogoliubov unitaries $\mathcal{B}_{\nu}$ are decomposed into elementary Gaussian unitaries\textemdash beam-splitters, displacement operators, and single-mode squeezing. Note that the displacement unitary for the $\nu$th fragment is denoted $\mathtt{D}^{(\nu)}$, which is distinct from the $\nu$th diagonal fragment $D_{\nu}$.  Each of these unitaries, along with the diagonal fragment unitaries $e^{-iD_{\nu}t}$, are then mapped to qubit operators using the unary boson-to-qubit mapping (Appendix~\ref{app:unary_map}), with the $l$-th mode truncated to have a maximum occupation number of $N_l-1$. The diagonal fragment unitaries are mapped to a product of $R_z$ rotations. Furthermore, the beam-splitter unitaries are mapped to exponents of commuting Pauli operators, which can also be simplified to products of $R_z$ rotations. On the other hand, the squeezing and displacement unitaries, when mapped to Pauli operators, result in exponents of non-commuting Pauli operators. These can be written as a product of Pauli exponents by introducing additional terms to ensure closure under commutation~\cite{izmaylov2020}. The full mapping procedure and simplification strategies are detailed in Appendix~\ref{app:bf_mapping}. As a result, the propagator for each fragment, and in turn the Trotterized propagator, can be decomposed into a product of Pauli rotations. As mentioned earlier, these Pauli rotations can be constructed using $R_z$ rotations and Clifford gates with the staircase algorithm \cite{Mansky2023}, while also having that the fully controlled time evolution can be implemented by adding a control on each of $R_z$ rotations coming from the diagonal fragment.

\begin{table*}
    \centering  
    \vspace{0.5cm}
    \renewcommand{\arraystretch}{1.7}  
    \begin{tabular}{|c|c|c|c|}
    \hline
    Form & Fragmentation Scheme & Number of qubits & Cost per Trotter oracle \\
    \hline 
    \multirow{2}{*}{Christiansen} & PF / FC / QWC & $MN$ & $\mathcal{O}\left( M^n N^{2n}\right)$\\
    & CGF & $MN$& $\mathcal{O}\left(M^n N^n N_{f,chr}\right)$ \\
    \hline
    Real Space & RS & $\left(M+\frac{d^2+3d-2}{2} \right) N_q + 2b_k + (d-1)b_r$& $\mathcal{O}\left( M^n (dN_q)^2 \right)$\\
    \hline
    Bosonic & BF & $MN$ & $\mathcal{O}\left(M^{\left\lceil\frac{d}{2}\right\rceil} N^{\left\lceil\frac{d}{2}\right\rceil} N_{f,bos}\right)$\\
    \hline
    \end{tabular}  
    \caption{Qubit counts and asymptotic T gate cost estimates for the various vibrational Hamiltonian forms considered in this work. $M$ is the number of vibrational modes, $N$ is the number of modals per mode, $N_q\sim\mathcal{O}(\log_2 N)$ is the number of qubits used per vibrational mode in the real space form, $n$ is the order of mode-couplings included in the potential, $d$ is the order of Taylor expansion for the potential, and $N_{f,chr/bos}$ is the number of fragments for Christiansen or bosonic forms. We have also used $b_k$ as the number of qubits for storing coefficients and $b_r$ for the number of qubits per resource register that are used for the quantum arithmetic-based implementation of the real space form in Fig.~\ref{fig:rs_exponential}. Refer to the text for more details.}
    \label{tbl:cost_estimates}
\end{table*}

\subsection{Trotter Error Analysis}
The Trotter error is estimated using Eq.~\eqref{eq:trial_state_trot_error}, where the contribution from the $n$th eigenstate is given in Eq.~\eqref{eq:nth_trot_error}. For the bosonic fragmentation scheme, the operator $\Theta_2$ is calculated for fast-forwardable bosonic fragments obtained using the algorithm mentioned in Sec.~\ref{sec:bosonic_frag}, with quartic fragments of the form given in Eq.~\eqref{eq:sol-bos-frags} and a quadratic fragment of the form of the last term in Eq.~\eqref{eq:diag_Hlq}. To calculate the Trotter error for selected molecules, the trial state is chosen as $\mu \ket{\Phi_0}$, where $\mu$ is the dipole moment operator in the \gls{HO} basis and $\ket{\Phi_0}$ denotes the harmonic ground state.

\subsection{Resource Estimates}
The $R_z$ gate count per Trotter oracle for the bosonic fragmentation scheme with a fourth-order Taylor expansion ($d=4$) is 
\begin{multline}
N_{R_z}^{\text{BF}} = 2M N  \\ +   \left[ \frac{M(M-1)}{2} N^2 + M \frac{ N (N-1)}{2}  
+  M N \right] (2N_f-1) \\ + 2 \big[ 4  M(M-1) N^2
+ M (N - 1) +  M N \big] (N_f+1)  ,
\end{multline}
where $N_f$ is the number of fragments. Additional $R_z$ gates result from the inclusion of compensating terms introduced to enforce closure in the displacement and squeezing unitaries. However, these contributions are relatively minor compared to the dominant costs and are therefore neglected in this estimate. Note that, as described in Sec.~\ref{sec:cform_resource}, to calculate the cost of implementing $L_{\mathrm{max}}$ Trotter steps, we have accounted for the cost reduction achieved by merging unitaries at the boundary of adjacent Trotter oracles. Details on how this estimate is calculated are presented in Appendix~\ref{app:bf_res_est}, along with generalizations to higher orders of Taylor expansion. As shown there, the asymptotic cost scales as
\begin{equation}
\mathcal{C}_{BF} \sim \mathcal{O}(M^{\left\lceil\frac{d}{2}\right\rceil} N^{\left\lceil\frac{d}{2}\right\rceil}N_f), 
\end{equation}
where $d$ is the truncation order for the Taylor expansion. Note that the bosonic fragmentation scheme does not rely on a particular order in the $n$-mode expansion of the Hamiltonian\textemdash for a $d$-th-order Taylor expansion, all couplings up to $d$-mode couplings are included in the potential. Thus, the scaling of $R_z$ gate count per Trotter oracle for the bosonic fragmentation is \textbf{independent} of the $n$-mode coupling order used in the Hamiltonian. 

\section{Results} \label{sec:results}

Having described in detail the different Hamiltonian representations, their fragmentations, and the implementation of the associated time evolution, we now show constant-factor resource estimates for time evolution in these representations. We also verify the fidelity of our time evolution implementation by performing statevector simulations to generate vibrational spectra. 

\subsection{Resource Estimates}

Asymptotic expressions of the T gate cost of a Trotter oracle for various forms of the vibrational Hamiltonian, derived in the previous sections, are summarized in Table~\ref{tbl:cost_estimates}. In the resource estimation, we consider the following fragmentation schemes:
\begin{enumerate}
    \item \glsentrylong{CGF} (\gls{CGF}) that are constructed from the Hamiltonian in the Christiansen form in a greedy fashion as described in Sec.~\ref{sec:cform_frag}, 
    \item \glsentrylong{PF} (\gls{PF}), \glsentrylong{FC} (\gls{FC}) fragments, and \glsentrylong{QWC} (\gls{QWC}) fragments for the Christiansen Hamiltonian mapped to Paulis, as described in Sec.~\ref{sec:cform_to_pauli}, 
    \item Fragments for the real space approach (RS) as described in Sec.~\ref{sec:rs_frag} 
    \item \glsentrylong{BF} (\gls{BF}) for the canonical bosonic form of the Hamiltonian, as described in Sec.~\ref{sec:bosonic_frag}.
\end{enumerate}
For the \gls{PF}, \gls{FC}, and \gls{QWC} fragments, the cost per Trotter oracle is the same irrespective of the fragmentation scheme, whereas the number of Trotter steps $r$ will be determined by the chosen fragmentation scheme. For both Christiansen and bosonic forms, the scaling of the number of fragments $N_f$ with respect to the system size ($M$ and $N$) is not known presently, which is why we report the scaling with the $N_f$ prefactor. For the real space approach, the number of qubits scales as $N_q ~\sim O(\log_2N)$ \cite{macridin2018a,macridin2018b}. 

While the asymptotic resource requirement expressions already allow to draw certain conclusions regarding the relative strengths and weaknesses of different Hamiltonian representations, these comparative advantages will be modified once Trotter error is taken into account. More specifically, the total resource estimate is affected by the number of Trotter steps required to achieve fixed error. The dependence of the number of Trotter steps on the system size is not known \textit{a priori}, thus necessitating empirical calculation of cost estimates for molecules of interest. 

In our constant factor resource estimation, we focus on four molecules: H$_2$S, CO$_2$, CH$_2$O, and CH$_4$. More details about these molecules are presented in Table~\ref{tbl:mols}. The Hamiltonians for these molecules are generated using the open-source software library PennyLane \cite{pennylane}. The Christiansen Hamiltonian is generated using \texttt{qchem.christiansen\_hamiltonian}, both in the harmonic oscillator modal basis and the mean-field optimized \gls{VSCF} modal basis. The bosonic Hamiltonian is generated using \texttt{qchem.taylor\_hamiltonian}, whereas the integrals for the real space Hamiltonian are generated by \texttt{qchem.taylor\_coeffs}. In all cases, we use restricted Hartree-Fock (RHF) as the electronic structure method to generate the underlying \gls{PES}. While using RHF likely gives an oversimplified description of the \gls{PES}, the choice of the electronic structure solver has negligible impact on resource estimates: for this reason we pick the solver that can generate the \gls{PES} quickly. Finally, we also use PennyLane to partition the qubit Hamiltonian into \gls{FC} or \gls{QWC} fragments using the largest first method. 

\begin{table}
    \centering  
    \vspace{0.5cm}
    \renewcommand{\arraystretch}{1.3} 
    \begin{tabular}{|c|c|c|}
    \hline
    \textbf{Molecule} & \textbf{Number} & \textbf{Number}  \\
     & \textbf{of} & \textbf{of modals}  \\
     & \textbf{modes} $\mathbf{(M)}$ & \textbf{per mode} $\mathbf{(N)}$ \\
    \hline 
    H$_2$S  & 3 & 4 \\ 
    CO$_2$  & 4 & 4 \\
    CH$_2$O & 6 & 4 \\
    CH$_4$  & 9 & 3 \text{ or } 4\\
    \hline
    \end{tabular}  
    \caption{Details about the various molecules considered in this work. The number of modals chosen is for cost estimation calculations in the bosonic and Christiansen forms. We have chosen $N_q = 4$ for the real space form.}
    \label{tbl:mols}
\end{table}

For all fragmentation schemes, we consider 2M4T Hamiltonians constructed in either normal mode coordinates or local mode coordinates, as described in Sec.~\ref{sec:setup}. 2M4T Hamiltonians include up to $2$-mode couplings, and further invoke a Taylor expansion of the mode coupling terms to $4$th order. Using the same 2M4T Hamiltonian facilitates a fair comparison of the different forms, although in the case of the Christiansen form one could use the full $2$-mode \gls{PES} at exactly the same cost, which would correspond to a so-called 2M\raisebox{0.25ex}{\small$\infty$}T Hamiltonian. This fact, alongside the capacity of working directly in the \gls{VSCF} basis, distinguishes the Christiansen form. For completeness, we also provide cost estimates for the Christiansen form using the \gls{VSCF} basis on the 2M\raisebox{0.25ex}{\small$\infty$}T Hamiltonian in the resource estimate table that follows, in parentheses. Note that using the \gls{VSCF} basis also allows for using a smaller number of modals which will entail a significant cost reduction, although for simplicity this reduction is not considered in our results.

\begin{figure}[t]
    \centering
    \includegraphics[width=0.95\columnwidth]{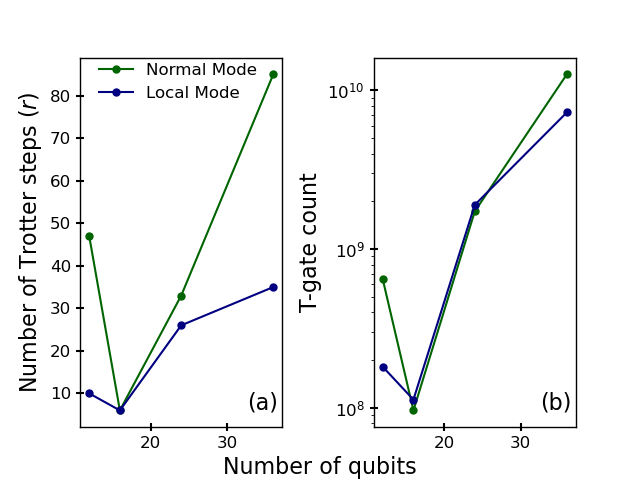}      
    \caption{(a) Number of Trotter steps $(r)$, and (b) T gate count for the \gls{PF} are compared for 2M4T normal-mode and local-mode Hamiltonians in \gls{HO} basis.}
    \label{fig:ham_comp}
\end{figure}

\begin{table*}[]
    \centering
    \renewcommand{\arraystretch}{1.5} 
    \begin{tabular}{|c|c|c|c|c|c|c|c|} \hline
    \textbf{Molecule} & \textbf{Form} & \textbf{Fragmentation} & \textbf{Qubits}  & \textbf{Fragments} &\textbf{Maximum } & \textbf{Trotter} $r$ & \textbf{T gates} \\ 
      &  &  \textbf{Scheme} & & & \textbf{T gates} & & \textbf{per Trotter call} \\ \hline
    \multirow{6}{*}{$\rm H_2S$} & \multirow{4}{*}{Chrisitiansen} & \gls{PF} & $12$ & $641 \, (718)$ &  $ 1.82 \, (0.82) \times 10^8$& $10 \, (4)$& $6.07 \,(6.80) \times 10^4$ \\ \arrayrulecolor{gray!50}\cline{3-8}\arrayrulecolor{black}
     & & \gls{QWC} & $12$ & $13 \, (12)$ & $ 4.30 \,(0.93) \times 10^8$& $ 26 \, (5)$& $ 6.00 \, (6.71) \times 10^4$ \\ \arrayrulecolor{gray!50}\cline{3-8}\arrayrulecolor{black}
     & & \gls{FC} & $12$ & $25\, (27)$ & $13.9 \, (1.95) \times 10^7 $& $8 \, (1)$& $6.03 \,(6.76) \times 10^4$\\ \arrayrulecolor{gray!50}\cline{3-8}\arrayrulecolor{black}
    &  & \gls{CGF} & $12$ & $3 \, (6)$ & $\mathbf{9.55} \,  (13.0) \mathbf{\times 10^6} $& $\mathbf{2}  \, (1)$& $ \mathbf{1.71} \,(4.43) \mathbf{\times 10^4}$\\ \arrayrulecolor{gray!50}\cline{2-8}\arrayrulecolor{black}
    & Bosonic & \gls{BF} & $12$ &  $3$ &$3.03 \times 10^8$ & $10$ & $1.29 \times 10^5$ \\ \arrayrulecolor{gray!50}\cline{2-8}\arrayrulecolor{black}
    & Real space & RS & 110 & $2$ & $4.02\times 10^8$ & $16$ & $8.38\times 10^4$\\ \hline 
    \multirow{6}{*}{$\rm CO_2$} & \multirow{4}{*}{Christiansen} & \gls{PF} & $16$ & $664\, (711)$ & $1.13 \, (1.21) \times 10^8$& $6 \, (6)$& $6.29 \,(6.73) \times 10^4$\\ \arrayrulecolor{gray!50}\cline{3-8}\arrayrulecolor{black}
     & & \gls{QWC} & $16$ &  $15 \, (16)$ &$2.58 \,(1.65) \times 10^8$& $15 \, (9)$& $6.10 \,(6.53) \times 10^4$\\ \arrayrulecolor{gray!50}\cline{3-8}\arrayrulecolor{black}
     & & \gls{FC} & $16$ & $15 \, (16)$ &  $5.46 \, (5.85 )\times 10^7 $& $\mathbf{3} \, (3)$& $6.21 \,(6.64) \times 10^4$\\ \arrayrulecolor{gray!50}\cline{3-8}\arrayrulecolor{black}
        & & \gls{CGF} & $16$ & $3 \, (3)$ & $ \mathbf{2.32} \, (2.32) \mathbf{\times 10^7} $& $\mathbf{3} \, (3)$& $ \mathbf{2.58} \, (2.58) \mathbf{\times 10^4}$\\ \arrayrulecolor{gray!50}\cline{2-8}\arrayrulecolor{black}
    & Bosonic & \gls{BF} & $16$ &  $3$ &$5.72 \times 10^8$ & $10$ & $2.43 \times 10^5$ \\ \arrayrulecolor{gray!50}\cline{2-8}\arrayrulecolor{black}
    & Real space & RS & 123 & $2$ & $2.27\times 10^8$ & $12$ & $6.3\times 10^4$ \\ \hline    
    \multirow{6}{*}{$\rm CH_2O$} & \multirow{4}{*}{Christiansen} & \gls{PF} & $24$ & $2568 \, (2840)$ & $ 1.90 \, (1.29) \times 10^9$& $26  \, (16)$& $2.43 \, (2.69) \times 10^5$\\ \arrayrulecolor{gray!50}\cline{3-8}\arrayrulecolor{black}
     & & \gls{QWC} & $24$ & $17 \, (20)$ & $ 2.10 \,(1.69) \times 10^9$& $ 32 \,(23)$& $2.35 \,(2.64) \times 10^5$\\ \arrayrulecolor{gray!50}\cline{3-8}\arrayrulecolor{black}
     & & \gls{FC} & $24$ & $35 \, (36)$ &  $4.88 \, (5.45) \times 10^8 $ & $7 \, (7)$ & $2.43 \,(2.67) \times 10^5$\\ \arrayrulecolor{gray!50}\cline{3-8}\arrayrulecolor{black}
    & & \gls{CGF} & $24$ & $5 \, (5)$ & $\mathbf{6.96} (10.4) \mathbf{\times 10^7} $ & $\mathbf{2} \, (3)$ & $\mathbf{1.11} \,(1.11) \mathbf{\times 10^5}$ \\ \arrayrulecolor{gray!50}\cline{2-8}\arrayrulecolor{black}
    & Bosonic & \gls{BF} & $24$ & $4$ &$2.63 \times 10^9$ &  $13$ & $7.99 \times 10^5$ \\ \arrayrulecolor{gray!50}\cline{2-8}\arrayrulecolor{black}
    & Real space & RS & 131 & $2$ & $1.57\times 10^9$ & $19$ & ${2.75\times 10^5}$ \\ \hline 
    \multirow{6}{*}{$\rm CH_4$} & \multirow{4}{*}{Christiansen} & \gls{PF} & $36$ & $7297 \, (7693)$ & $ 7.26 \, (3.94) \times 10^9$& $ 35 \, (18)$& $ 6.91 \,(7.29) \times 10^5$\\ \arrayrulecolor{gray!50}\cline{3-8}\arrayrulecolor{black}
     & & \gls{QWC} & $36$ & $24 \, (24)$ & $11.9 \,(4.6) \times 10^{9}$& $ 63 \, (23)$& $ 6.79 \,(7.12) \times 10^5$\\ \arrayrulecolor{gray!50}\cline{3-8}\arrayrulecolor{black}
     & & \gls{FC} & $36$ & $49 \, (57)$ & $22.0 \, (6.39) \times 10^8 $& $11 \, (3)$& $6.91 \, (7.27) \times 10^5$\\ \arrayrulecolor{gray!50}\cline{3-8}\arrayrulecolor{black}
    & & \gls{CGF} & $36$ & $5 \, (8)$ & $\mathbf{3.68} \, (2.70) \mathbf{\times 10^8}$& $\mathbf{5} \, (2)$ & $\mathbf{2.28} \,(4.33) \mathbf{\times 10^5}$ \\ \arrayrulecolor{gray!50}\cline{2-8}\arrayrulecolor{black}
    & Bosonic & \gls{BF} & $27$ & $4$ & $3.83 \times 10^{9\mathbf{*}}$ & $18^{\mathbf{*}}$ & $8.43 \times 10^{5\mathbf{*}}$ \\ \arrayrulecolor{gray!50}\cline{2-8}\arrayrulecolor{black}
    & Real space & RS & 143 & $2$ & $5.72\times 10^9$ & $25$ & ${7.64\times 10^5}$ \\ \hline  
    \end{tabular}
    \caption{Resource estimates for local mode 2M4T Hamiltonians in HO basis for the various fragmentation schemes investigated in this work. Cheapest estimates for each molecule for the local mode 2M4T Hamiltonians are highlighted in bold. For the  Christiansen form, we also report in parenthesis, cost estimates for 2M\raisebox{0.25ex}{\small$\infty$}T Hamiltonians in a VSCF basis. Refer to the main text for details. $^{\mathbf{*}} $ indicates that the cost estimates for $\rm CH_4$ in the bosonic form are for $N=3$, where as all others are for $N=4$.}
    \label{tab:estimates}
\end{table*}

With the Hamiltonians obtained, we calculate the T gate cost for a fixed total evolution time $t_{\text{max}}$. The total evolution time is chosen to mimic the longest evolution time required for a vibrational spectrum calculation, as presented in Ref.~\cite{loaiza2025}. A full spectrum calculation would require varying number of shots for time evolutions of varying lengths, as detailed in Ref.~\cite{loaiza2025}: our cost estimate thus corresponds to the cost of the deepest circuit required in the vibrational spectrum calculation. The total evolution time $t_{\text{max}}$ is divided into $k_{\text{max}}$ steps of time $\tau$ each. We chose $\tau = 250~\text{a.u.} \approx 6$ fs, which corresponds to a spectral range of $2\pi/\tau \approx 5516{\rm \ cm}^{-1}$.  We choose the total evolution time of $k_{\text{max}} = 300$, which is typically sufficient to obtain a spectrum with a resolution of $10$ cm$^{-1}$ \cite{qpe_qmegs,loaiza2025}. 

The propagation for a time step $\tau$ is divided into $r$ Trotter steps. As a result, a total of $L_{\text{max}} = r \times k_{\text{max}}$ Trotter steps are required for the total evolution. The number of Trotter steps $r$ is calculated as described in Sec.~\ref{sec:trotter_error} with $\epsilon_{\text{trot}} = 7$ cm$^{-1}$. The total T gate cost is thus determined by two factors, the cost of implementing one Trotter oracle and the total number of calls to the Trotter oracle -- both of which are determined by the Hamiltonian form, the coordinates chosen, and the fragmentation scheme.

\begin{figure*}
    \centering
    \includegraphics[width=0.9\textwidth]{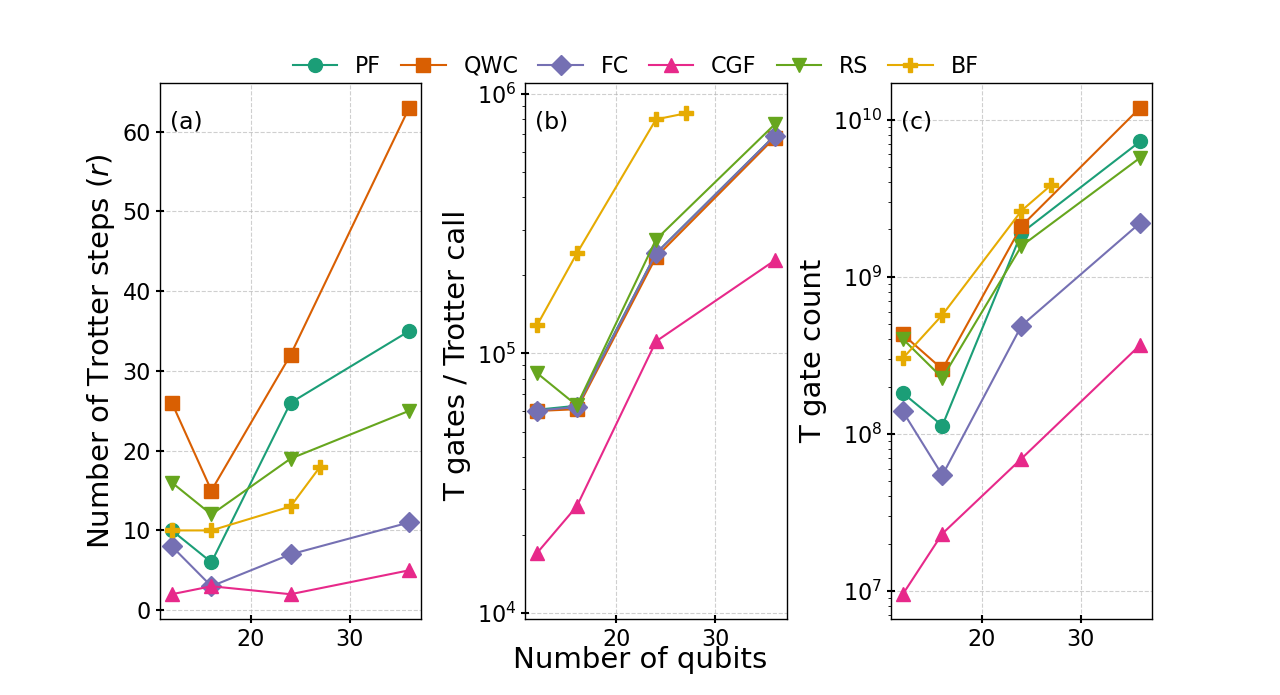}  
    \caption{Comparison of the cost estimates for 2M4T local-mode Hamiltonians in \gls{HO} basis for various fragmentation schemes: (a) number of Trotter steps, (b) T gates / Trotter call, (c) T gate count. Note that the plots for \gls{PF}, \gls{QWC}, and \gls{FC} schemes overlap in panel (b).}
    \label{fig:frag_comp}
\end{figure*}

We begin by investigating the effect of the choice of coordinates (normal versus local modes) on the T gate cost estimates. Across all the fragmentation methods studied, the cost of simulating local-mode Hamiltonians is observed to be similar to or lower than normal-mode Hamiltonians. As an illustration, Fig.~\ref{fig:ham_comp} presents the number of Trotter steps $r$ and the T gate cost for the \gls{PF} method. As highlighted in Fig.~\ref{fig:ham_comp}(a), consistent with observations for other fragmentation schemes, the primary source of reduction in cost is the smaller number of Trotter steps $r$ for local mode Hamiltonians as compared to normal mode Hamiltonians: the other factor determining the total T gate cost, namely the cost per Trotter oracle, is largely unaffected by changing to the local mode form. While it seems reasonable that a more localized mode basis might reduce overall inter-fragment non-commutativity and thus Trotter error, further investigations are needed to understand exactly why this occurs. In addition to the smaller number of Trotter steps observed here for local mode Hamiltonians, for a given order of mode couplings, local mode Hamiltonians are expected to be a more accurate description of the PES than normal mode Hamiltonians \cite{mode_loc_1}, as we discussed earlier in Sec. \ref{sec:setup}. Hence, from here on, we focus on local mode Hamiltonians for the comparison of costs across fragmentation schemes, although similar results were obtained for the normal mode case.

Costs for the different fragmentation schemes are presented in Table~\ref{tab:estimates} and also plotted in Fig.~\ref{fig:frag_comp} to facilitate comparison. For the Christiansen form, Table~\ref{tab:estimates} also provides, in brackets, cost estimates for the 2M\raisebox{0.25ex}{\small$\infty$}T Hamiltonians in the \gls{VSCF} basis, which provide an improved description of the vibrational problem compared to the 2M4T Hamiltonians in the \gls{HO} basis. The T gate counts are split into their two constituent components -- the number of T gates per call to the Trotter oracle and the number of calls to the Trotter oracle. For fixed $k_{\text{max}}$ considered here, the latter is determined by the number of Trotter steps $r$. 

First, we compare the costs for the different fragmentation schemes for the Christiansen Hamiltonian mapped to Paulis. For these schemes, the T gate cost for each Trotter oracle is very similar, as discussed in Sec.~\ref{sec:cform_resource}. Therefore, the variation in T gate count across these is only determined by the number of Trotter steps $r$. Figure~\ref{fig:frag_comp}(a) shows that, as expected, the number of Trotter steps is smaller for the \gls{FC} scheme than the naive \gls{PF} scheme that treats each Pauli as an individual fragment (no grouping). However, surprisingly, the number of Trotter steps for the \gls{QWC} scheme is larger than for the naive \gls{PF} scheme. This highlights that the number of Trotter steps depends intricately on the commutation relations between the fragments, and not just on the number of fragments. As a result, the \gls{FC} scheme is the cheapest of the three fragmentation schemes for the Christiansen Hamiltonian mapped to Paulis. 

Comparing between forms, we note that the BF scheme has a number of Trotter steps comparable to the other schemes, but also a very high T gate cost per Trotter oracle -- making it the most expensive of the schemes studied here. The \gls{RS} approach has both the number of Trotter steps and cost of each Trotter oracle comparable to the \gls{PF} scheme, making it overall comparable in cost to the \gls{PF} scheme.  
Interestingly, the real space form has a moderately large number of Trotter steps despite having only two fragments, highlighting again that the Trotter errors do not depend only on the number of fragments but have a complex dependence on the commutation relations between the fragments. 

\begin{figure*}
    \centering
    \includegraphics[width=0.95\textwidth]{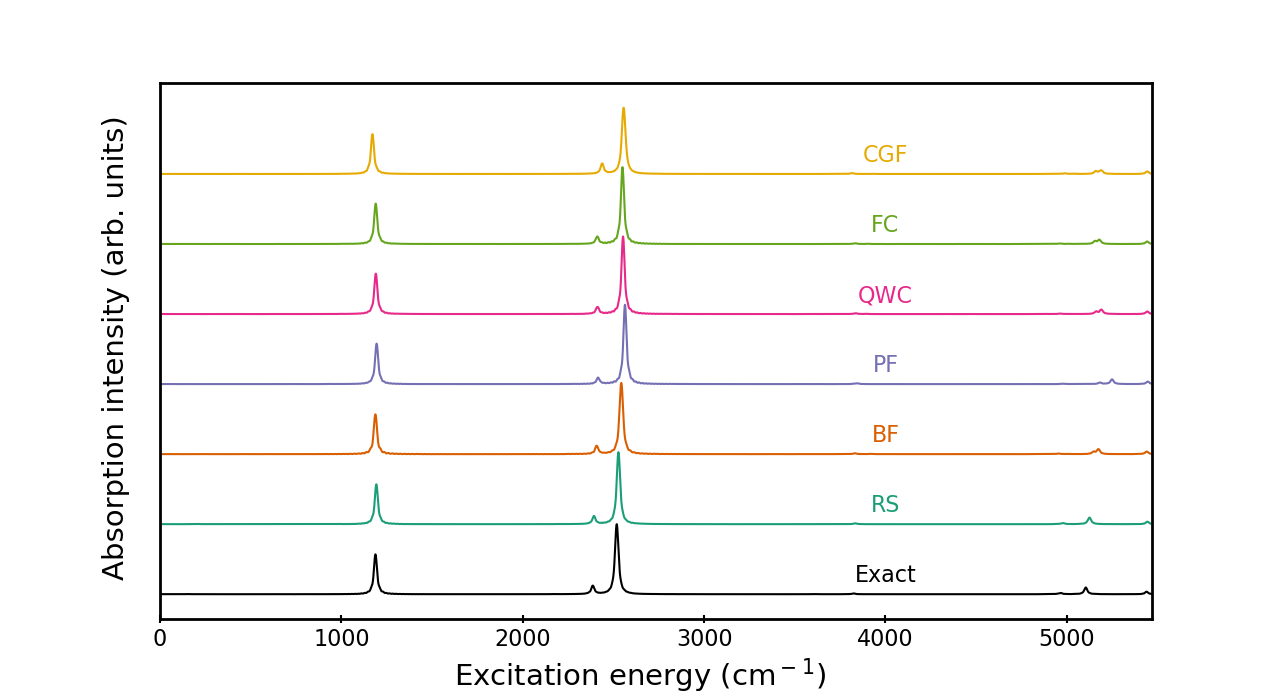}
    \caption{The IR spectrum for H$_2$S calculated exactly and using the Trotter approximation for various fragmentation schemes. The spectra for the fragmentation schemes are in excellent agreement with the exact spectrum, demonstrating the accuracy of our framework. For the real space approach, each mode is discretized using $N_q = 4$ qubits, whereas for the bosonic and Christiansen based approaches, $N = 4$ modals are used for each mode.}
    \label{fig:spec_frag_comp}
\end{figure*}

Lastly, the \gls{CGF} scheme for the Christiansen form performs well on both fronts. It has the smallest number of Trotter steps and the cheapest T gate cost per Trotter oracle, making it the most efficient fragmentation scheme studied here. Compared to the prior-art of the real space approach \cite{loaiza2025} the results in Table \ref{tab:estimates} clearly show a factor of 10 speedup using the \gls{CGF} scheme. While we did not explicitly study the azidoacetylene molecule and other larger systems from Ref.~\cite{loaiza2025}, we expect the comparison to be the same, as the Christiansen and real space forms have the same scaling of $M^n$ with the number of modes $M$.  In addition to that, for the Christiansen form Table~\ref{tab:estimates} also provides cost estimates for the 2M\raisebox{0.25ex}{\small$\infty$}T Hamiltonians in the \gls{VSCF} basis, which provide an improved description of the vibrational problem than the 2M4T Hamiltonians in the \gls{HO} basis. The data presented show that this improved description can be obtained at a comparable cost for the \gls{CGF} scheme, and at a comparable or slightly lower cost for the Pauli-based fragmentation schemes. The slight improvements in cost are largely due to the use of the optimized VSCF basis.  By contrast, such a description would be a factor of $n^2 / d^2 = 16^2 / 4^2 = 16$ more expensive (for a \gls{PES} computed over 16 grid points) based on the asymptotic scaling for the real space form, and require roughly 10 times as many qubits to store the results of the intermediate multiplications. The costs would grow even more quickly in the bosonic form where the cost scales as $M^d$ with the number of modes.

\subsection{Spectra}

As a final benchmark, we use all of the fragmentation schemes and their associated time evolution implementations we described to calculate the infrared absorption spectrum for the H$_2$S molecule using the \gls{FT} of the autocorrelation function for the normalized trial state $\ket{\psi} \propto \mu \ket{\Phi_0}$, defined as, 
\begin{align}
    C(t) = \mel{\psi}{e^{-iHt}}{\psi},
\end{align}
where $\mu$ is the dipole operator and $\ket{\Phi_0}$ is the \gls{HO} ground state. The time evolution is implemented exactly as presented in the previous sections, using the PennyLane \texttt{lightning} \cite{asadi2024} statevector simulator. The total evolution time is divided into $k_{\text{max}}$ steps of time $\tau$ each. Here we chose $\tau = 250~\text{a.u.} \approx 6$ fs, which corresponds to a spectral range of $2\pi/\tau \approx 5516{\rm \ cm}^{-1}$. We choose $k_{\text{max}} = 300$, which is sufficient to obtain a spectrum with an accuracy of $10$ cm$^{-1}$ when a matching pursuit algorithm is used \cite{qpe_qmegs,loaiza2025}. We also included a Lorentzian broadening of $\eta = 10$ cm$^{-1}$ while performing the \gls{FT}. The spectrum thus obtained yields peaks at the eigenenergies of the Hamiltonian, and thus requires setting the \gls{ZPE} to zero to obtain the peaks at the transition frequencies. We approximate the \gls{ZPE} here as $E_0 \approx \mel{\phi_0}{H}{\phi_0}$ to set the zero of energy in the spectrum. As the primary goal of the spectrum calculation here is to demonstrate the reliability of the proposed fragmentations and the perturbative Trotter time step estimates,  this energy shift does not affect the analysis.

\begingroup
\setlength{\tabcolsep}{5pt} 

\begin{table*}[t]
    \centering  
    \renewcommand{\arraystretch}{1.3} 
    \begin{tabular}{|p{2.6cm}|p{3.5cm}|p{3.5cm}|p{3cm}|p{3.5cm}|}
    \hline
    \textbf{Representation} & \textbf{Strength} & \textbf{Reasoning} & \textbf{Weakness} & \textbf{Reasoning}  \\
    \hline 
    \raisebox{-7.5ex}{Christiansen} & 
    \begin{itemize}[itemsep=0.5ex, leftmargin=*]
        \item \textbf{Great general purpose}
        \item When PES is complex
    \end{itemize} & 
    \begin{itemize}[itemsep=0.5ex, leftmargin=*]
        \item Cheapest Trotter step and lowest error
        \item Generic PES at no extra cost
    \end{itemize} & 
    \begin{itemize}[itemsep=0.5ex, leftmargin=*]
        \item High-mode coupling
        \item Large number of modals
    \end{itemize} & 
    \begin{itemize}[itemsep=0.5ex, leftmargin=*]
        \item Cost grows quickly with modals
        \item Cost grows quickly with mode couplings
    \end{itemize} \\ 
    \hline
    \raisebox{-3.5ex}{Bosonic} & 
    \begin{itemize}[itemsep=0.5ex, leftmargin=*]
        \item High-mode coupling but PES is simple
    \end{itemize} & 
    \begin{itemize}[itemsep=0.5ex, leftmargin=*]
        \item Cost is independent of mode coupling
    \end{itemize} & 
    \begin{itemize}[itemsep=0.5ex, leftmargin=*]
        \item Highest cost of all representations
    \end{itemize} & 
    \begin{itemize}[itemsep=0.5ex, leftmargin=*]
        \item High Trotter error and per-step cost
    \end{itemize} \\
    \hline
    \raisebox{-8.5ex}{Real space} & 
    \begin{itemize}[itemsep=0.5ex, leftmargin=*]
        \item High energy excitations requiring many modals
        \item High-frequency dynamics
    \end{itemize} & 
    \begin{itemize}[itemsep=0.5ex, leftmargin=*]
        \item Scaling with modals is logarithmic
        \item Shorter steps reduce Trotter error the most of all approaches
    \end{itemize} & 
    \begin{itemize}[itemsep=0.5ex, leftmargin=*]
        \item High qubit count
        \item High Trotter error
    \end{itemize} & 
    \begin{itemize}[itemsep=0.5ex, leftmargin=*]
        \item Large ancilla overhead for arithmetic
        \item Kinetic and potential terms have large non-commutativities
    \end{itemize} \\
    \hline
    \end{tabular}  
    \caption{Summary of the strengths and weaknesses of the three respective vibrational Hamiltonian representations studied in this work.}
    \label{tab:comparative_advantage}
\end{table*}

\endgroup

In Fig.~\ref{fig:spec_frag_comp} the spectrum obtained from exact diagonalization for the H$_2$S molecule is compared with those obtained from the Trotterized propagator using Trotter time steps from the perturbative time step estimation described in Sec.~\ref{sec:trotter_error}. The exact and real space spectrum are obtained by using a discretization of $N_q = 4$ along each vibrational mode, whereas the Christiansen and bosonic spectra are obtained by using $N = 4$ modals per mode. The small discrepancy seen in the peaks $\sim 2500$ cm$^{-1}$ and $\sim 5000$ cm$^{-1}$ in the Christiansen and bosonic spectra, when compared to the exact spectrum, are due to the finer discretization using $N_q = 4$ in the latter as compared to using $N = 4$ modals in the former. This has been verified by copmaring with the spectrum obtained by exact diagonalization of the Hamiltonian in the Christiansen form with $N=4$ modals. The overall excellent agreement between the spectra obtained using the Trotterized propagator in the different forms demonstrates the reliability of the fragmentation and time evolution implementations, as well as of the perturbative Trotter error estimation method to correctly predict the required time step without being too conservative.

\section{Comparative advantages of Hamiltonian representations} \label{sec:comp_forms}

With the asymptotic scalings and resource estimates laid out for all cases, we now address the question of when a given Hamiltonian representation should be used in practice. Our analysis allows us to delineate clear regimes of advantage for each representation, supported by data from Table~\ref{tbl:cost_estimates} and Table~\ref{tab:estimates}: 
\begin{itemize}
    \item \textbf{Bosonic:} One clear strength of the bosonic approach evident from the asymptotic expressions is that it only depends on the Taylor approximation order $d$, not on the order of the $n$-mode expansion. Thus, whenever the Hamiltonian couples more than two modes at a time, the bosonic simulation method will be highly competitive, as long as the Taylor order required to represent the PES is not too high. The T gates per Trotter call for other forms will become more expensive as higher mode couplings are included in the potential, but this will not be the case for the bosonic form. Thus, if the Trotter error behavior remains qualitatively similar to what we see in Table~\ref{tab:estimates} for higher mode couplings, the bosonic form may be comparable in cost, or even cheaper, than other forms.
    \item \textbf{Christiansen:} Conversely, if an accurate representation of the PES is paramount, then the Christiansen form is ideal as its cost is independent of the Taylor approximation degree $d$. This is shown explicitly in Table~\ref{tab:estimates} where the Christiansen form costs for 2M\raisebox{0.25ex}{\small$\infty$}T Hamiltonians are extremely similar as for 2M4T. By comparison, the bosonic form cost will quickly become severely expensive in this case (the cost of the real space form will rise too, but only polynomially).
    \item \textbf{Real space:} Finally, if access to higher excited states is sought, such that high-energy modals become relevant and need to be explicitly included in the simulation, the real space form has the best scaling for adding modals to the system representation. The accuracy of higher excited states is also controlled by the underlying potential though, so the need for a higher-order Taylor expansion in the real space form may offset its cheaper asymptotic scaling with respect to the number of modals. 
    
    Another regime where the real space form may have an advantage is in short-timestep (equivalently, high-frequency) dynamics. For a small $\tau$, the number of Trotter steps $r$ will decrease for all the forms. However, for approaches like Christiansen form that already have a small $r$, the reduction in cost will be relatively small (as it cannot go below $r=1$), as compared to the real space scheme, which could benefit from a significantly larger cost reduction. 
\end{itemize}

We summarize these conclusions succinctly in Table~\ref{tab:comparative_advantage}.

\glsresetall
\section{Conclusions} \label{sec:conc}

Quantum algorithms are a promising alternative to classical simulations of vibrational spectroscopy and dynamics. In this work, we have presented highly optimized quantum algorithms for performing Trotterized time evolution of vibrational Hamiltonians that have achieved a factor of 10 speedup over prior-art algorithms, making the use of quantum algorithms for vibrational dynamics an attractive use case for fault tolerant quantum computers. We have studied three distinct realizations of the vibrational Hamiltonian in great detail: the canonical bosonic second quantization form, the Christiansen second quantized form, and the real space form. This included developing new fragmentation schemes using the generalization of the Cartan subalgebra (CSA) approach, and combining them with a perturbative estimate of Trotter time steps to obtain T gate costs for simulating time evolution, using quantum circuits for each form. Overall, we found that the Christiansen form with the \gls{CGF} fragmentation scheme is the most attractive approach for most vibrational Hamiltonian simulation contexts, with the real space and bosonic approaches providing competitive alternatives in cases of large number of modals or high-mode coupling, respectively. 

Looking ahead, we anticipate that more sophisticated fragmentation schemes, as well as exploiting symmetries, could be leveraged to further optimize costs of the Christiansen form \cite{martinez2023, mehendale2025}. On the bosonic front, a more efficient implementation of Gaussian gates using elementary qubit gate sets could lead to cost reductions. For the \gls{RS} approach, improvements to arithmetic subroutines are the likely source of further optimization. Furthermore, the fragmentation schemes introduced in this work, especially for the Christiansen form, could also be used in approaches based on qubitization. Overall, there is significant scope for further improvement of quantum algorithms for vibrational dynamics. We hope the extensive study of the possible Hamiltonian representations from a unified Lie-algebraic perspective presented here can serve as a useful foundation for continuing to develop vibrational dynamics into an attractive application for quantum computing.

\section*{Acknowledgements}
S.M. thanks Shashank Mehendale, Smik Patel, and Paarth Jain for insightful discussions. S.K. acknowledges Smik Patel for valuable inputs on the bosonic fragmentation scheme. This work was funded under the NSERC Quantum Alliance program. This research was partly enabled by Compute Ontario (computeontario.ca) and the Digital Research Alliance of Canada (alliancecan.ca) support. Part of the computations were performed on the Niagara supercomputer at the SciNet HPC Consortium. SciNet is funded by Innovation, Science, and Economic Development Canada, the Digital Research Alliance of Canada, the Ontario Research Fund: Research Excellence, and the University of Toronto. This research used resources of the National Energy Research Scientific Computing Center (NERSC), a Department of Energy User Facility using NERSC award DDR QIS-ERCAP ERCAP0032729.

\bibliography{bibfile}

\appendix 

\section{Unary Boson-to-Qubit Mapping} \label{app:unary_map}

Let us consider a bosonic system with $M$ modes. To map it to qubits, we need to truncate the maximum allowed occupation number. For the $l$th mode the maximum occupation number is chosen to be $N_l - 1$, resulting in a total of $N_l$ `modals' per mode. Then, in the unary mapping, the $l$th mode is mapped to $N_l$ qubits, resulting in a $N_{Qb} = \sum_{l=1}^{M} N_l$ qubit system. A general $M$-mode bosonic fock state, with $N_l$ modals for the $l$th mode, gets mapped to a $N_{Qb}-$qubit state as,
\begin{align}
    \ket{n_1,n_2,...,n_M} &= \otimes_{l=1}^{M}\left[\left(\otimes_{\alpha_l=0}^{n_l-1}\ket{0}_{\alpha_l}\right)\otimes\ket{1}_{n_l} \right. \notag \\
    &\left.\left(\otimes_{\alpha_l=n_l+1}^{N_l-1}\ket{0}_{\alpha_l}\right)\right].
\end{align}
Correspondingly, the bosonic ladder operators get mapped as, 
\begin{align}
    b_l^{\dagger} &= \sum_{\alpha_l = 0}^{N_l-2} \sqrt{\alpha_l + 1} \ket{0}\bra{1}_{\alpha_l} \otimes \ket{1}\bra{0}_{\alpha_l+1}, \\
     b_l &= \sum_{\alpha_l = 1}^{N_l-1} \sqrt{\alpha_l} \ket{1}\bra{0}_{\alpha_l-1} \otimes \ket{0}\bra{1}_{\alpha_l} .    
\end{align}

\section{Christiansen Greedy Fragmentation for the Three-mode Hamiltonian} \label{app:cform_three_mode}

The \gls{CGF} procedure can be extended to Hamiltonians with higher-order mode couplings. Here we illustrate it for the three-mode coupling case. $H = H_{1M} + H_{2M} + H_{3M}$ where, $ H_{1M}$ and $H_{2M}$ are defined in Eq.~\eqref{eq:c_form}, and,
\begin{align}
    H_{3M} = \sum_{l > m > n}\sum_{\substack{\alpha_l \beta_l \\ \gamma_m \delta_m \\ \zeta_n \eta_n}} f^{(lmn)}_{\alpha_l \beta_l \gamma_m \delta_m \zeta_n \eta_n} E^{\alpha_l}_{\beta_l} E^{\gamma_m}_{\delta_m} E^{\zeta_n}_{\eta_n}. 
\end{align}
The one mode term will also be a fast-forwardable fragment in this case, and the form of fast-forwardable fragments for the higher order terms is,
\begin{align}
    H_{\nu} & = \mathcal{G}_{\nu} \left(\sum_{l > m} \sum_{i_l j_m} \lambda^{(\nu, lm)}_{i_l j_m} n_{i_l}n_{j_m} \right. \notag \\
    & \left. + \sum_{l > m > n} \sum_{i_l  j_m  k_n} \xi^{(\nu,lmn)}_{i_l j_m k_n} n_{i_l}n_{j_m}n_{k_n} \right) {\mathcal{G}_{\nu}}^{\dagger}, 
\end{align}
The three-mode tensor of each fragment $\xi^{(\nu,lmn)}_{i_l j_m k_n}$ satisfies the same eight-fold symmetries as the three-mode tensor of the Hamiltonian. The GFRO algorithm presented in Sec.~\ref{sec:cform_frag} can be modified to find such fragments. For this purpose, the cost function is chosen to be the sum of $L_2$ norms of the two-mode and three-mode tensors. \\

\section{Decomposing Propagators for Christiansen Greedy Fragments into Pauli Rotations} \label{app:cf_mapping}

We decompose the fragment propagator for \glsentrylong{CGF} in Eq.~\eqref{eq:prop_cf} mapped to Pauli operators, into a product of Pauli rotations, to be able to implement it as a quantum circuit. For this purpose, both the modal rotation unitaries $\mathcal{G}_{\nu}$, and the diagonal fragment unitaries $e^{-iD_{\nu}t}$ for $\nu \in [0,N_f]$, need to be expressed as a product of Pauli rotations. Similar to the fermionic case \cite{kivlichan2018,kivlichan2020}, the  unitary $\mathcal{G}^{(l)}_{\nu}$ for each mode can be decomposed into rotations between modal pairs \cite{reck1994}, which can in turn be mapped to exponents of Pauli operators,
\begin{align}
    \mathcal{G}_{\nu} &= \prod_{l} \mathcal{G}^{(l)}_{\nu} = \prod_{l}e^{\sum_{\alpha_l>\beta_l}\theta^{(\nu,l)}_{\alpha_l\beta_l}\left(E^{\alpha_l}_{\beta_l} - E^{\beta_l}_{\alpha_l}\right)} \notag \\
    & = \prod_{l} \prod_{\alpha_l>\beta_l} e^{\phi^{(\nu,l)}_{\alpha_l\beta_l}\left(E^{\alpha_l}_{\beta_l} - E^{\beta_l}_{\alpha_l}\right)} \notag \\ 
    & =\prod_{l}\prod_{\alpha_l > \beta_l} e^{\frac{i}{2}\phi^{(\nu,l)}_{\alpha_l\beta_l}\left(x_{\alpha_l} y_{\beta_l} - x_{\beta_l} y_{\alpha_l}\right)}.  \label{eq:givens_rots}
\end{align}
Note that the parameters $\phi^{(\nu,l)}_{\alpha_l\beta_l}$ can be obtained from $\theta^{(\nu,l)}_{\alpha_l\beta_l}$. Similarly, the unitary corresponding to the diagonal one-mode fragment can be mapped to Pauli operators as,
\begin{align}
    e^{-iD_{0}t} &= e^{-i\left(\sum_{l} \sum_{\alpha_l} \epsilon^{(l)}_{\alpha_l} n_{\alpha_l}\right) t} = \prod_{l} \prod_{\alpha_l}e^{-i \epsilon^{(l)}_{\alpha_l} n_{\alpha_l}t} \notag \\
    & = \prod_{l} \prod_{\alpha_l}e^{-\frac{i}{2} \epsilon^{(l)}_{\alpha_l} \left(\bm{1}-z_{\alpha_l}\right)t} \label{eq:om_frag_unit}
\end{align}
and that for the diagonal two-mode fragments as, 
\begin{align}
    e^{-iD_{\nu}t} &= e^{-i\left(\sum_{l > m} \sum_{i_l,j_m} \lambda^{(\nu, lm)}_{i_l,j_m} n_{i_l}n_{j_m}\right) t} \notag \\
    & = \prod_{l > m} \prod_{i_l,j_m} e^{-i \lambda^{(\nu, lm)}_{i_l,j_m} n_{i_l}n_{j_m} t} \notag \\
    & = \prod_{l > m} \prod_{i_l,j_m} e^{-\frac{i}{4}\lambda^{(\nu, lm)}_{i_l,j_m} \left(\bm{1}-z_{i_l}-z_{j_m}+z_{i_l}z_{j_m}\right)t}. \label{eq:tm_frag_unit}
\end{align}
As a result, we have decomposed the propagator for each fragment unitary as a product of Pauli rotations.

\section{Resource Estimates for \glsentrylong{CGF} Scheme} \label{app:cf_res_est}

For these estimates, we assume that the number of modals for each mode is the same, $N_l = N$ for all $l \in [1,M]$, although the general case will follow a similar reasoning. Let us begin with the two-mode Hamiltonian, $n=2$. Consider that we have $N_f$ two-mode fragments and one one-mode fragment. The ordering of the fragments for the second-order Trotter propagator is chosen such that the one-mode fragment comes first, and is followed by the 2-mode fragments in the order they are found in the GFRO algorithm in Sec.~\ref{sec:cform_frag}. The $R_z$ gate count for a second-order Trotter propagator can be broken down into three different contributions.
\begin{enumerate}
    \item  From the structure of the second-order Trotter propagator in Eq.~\eqref{eq:trot2}, we note that $ \mathcal{G}^{\dagger}_{N_f}\mathcal{G}_{N_f} = \bm{1}$. Furthermore,  the unitaries $ \mathcal{G}_{\nu}$ form a group, allowing us to employ  $\mathcal{G}^{\dagger}_{\nu}\mathcal{G}_{\nu + 1} =  \mathcal{G}^{'}$ is also a modal rotation unitary. These simplifications result in a total of $2\left(N_f + 1\right)$ such unitaries. Following Eq.~\eqref{eq:givens_rots}, each unitary $ \mathcal{G}_{\nu}$ requires $M N(N-1)$ $R_z$ gates.

    \item There are a total of $2N_f$ two-mode diagonal unitaries. However, noting the simplification $e^{-iD_{\nu}\Delta t /2} \mathcal{G}^{\dagger}_{N_f}\mathcal{G}_{N_f} e^{-iD_{\nu}\Delta t /2} = e^{-iD_{\nu}\Delta t}$, results in $2N_f - 1$ distinct two-mode diagonal unitaries that need to be implemented. As shown in Eq.~\eqref{eq:tm_frag_unit}, each such unitary corresponds to $MN\left( 1 + \frac{1}{2}N(M-1)\right)$ $R_z$ gates.

    \item The two one-mode diagonal fragment unitaries each contribute $NM$ $R_z$ gates, as shown in Eq.~\eqref{eq:om_frag_unit}.
\end{enumerate}
Combining all contributions, the $R_z$ gate cost for each second-order Trotter oracle is,
\begin{multline}
N_{R_z}^{CGF}(\text{Trotter oracle}) = 2MN + 2MN(N-1)(N_f+1)  \\
+ MN\left( 1 + \frac{1}{2}N(M-1)\right)\left(2N_f -1 \right).
\end{multline}
The  $R_z$ cost of a long-time evolution with $L^{CGF}_{\text{max}}$ Trotter steps can be calculated similarly as,
\begin{multline}
    N_{R_z}^{CGF}(\text{Long time})  = \left(L^{CGF}_{\text{max}} + 1\right)MN  \\
+ 2MN(N-1)(N_f L^{CGF}_{\text{max}}+1)  \\
 + MN\left( 1 + \frac{1}{2}N(M-1)\right)\left(2N_f -1 \right)L^{CGF}_{\text{max}}.
\end{multline}
For a three-mode coupling Hamiltonian, $n=3$, the number of $R_z$ gates for each three-mode fragment unitary will be $MN\left( 1 + \frac{1}{2}N(M-1) + \frac{1}{6}N^2(M-1)(M-2)\right)$, while the other contributions remain the same as in the two-mode coupling case. This suggests that as the mode-coupling $n$ is increased, the dominant cost comes from the diagonal $n-$mode fragment unitaries. Thus, the $R_z$ gate cost per Trotter oracle for the $n-$mode coupling case scales as $\mathcal{O}\left(N_f M^n N^n\right)$.

\section{Resource Estimates for Fragments of Christiansen Hamiltonian Mapped to Paulis} \label{app:pauli_est}
The Hamiltonian for the \gls{PF} scheme, as mentioned in Sec.~\ref{sec:cform_frag}, is $H = \sum_{\nu=1}^{N_H} P_{\nu}$, where we have absorbed the coefficients of each term into the Pauli operator. The second-order Trotter propagator for the \gls{PF} scheme is then,
\begin{multline}
     e^{-iH\Delta t} \approx \prod_{\nu=1}^{N_H} e^{-iP_{\nu}\Delta t/2} \prod_{\nu=N_H}^{1} e^{-iP_{\nu} \Delta t/2} \\
    = \left(\prod_{\nu=1}^{N_H - 1} e^{-iP_{\nu}\Delta t/2}\right) e^{-iP_{N_H}\Delta t} \left( \prod_{\nu=N_H-1}^{1} e^{-iP_{\nu} \Delta t/2}\right), \notag 
\end{multline}
resulting in an $R_z$ cost per Trotter oracle for the \gls{PF} scheme of,
\begin{align}
    N_{R_z}^{PF}(\text{Trotter oracle}) = 2N_H - 1,     
\end{align}
Similarly, noting that the Pauli rotations at the end of one Trotter oracle and the beginning of the next can also be combined, the $R_z$ cost for a long-time evolution with $L^{PF}_{\text{max}}$ Trotter steps will be,
\begin{align}
    N_{R_z}^{PF}(\text{Long time}) = 2N_H L^{PF}_{\text{max}} - \left(2 L^{PF}_{\text{max}} - 1\right).   
\end{align}
For both \gls{QWC} and \gls{FC} schemes, a similar argument to obtain reductions by combining the last fragment in the second-order Trotter propagator results in an $R_z$ cost per Trotter oracle of,
\begin{align}
    N_{R_z}^{QWC/FC}(\text{Trotter oracle}) = 2N_H - N^{QWC/FC}_{\text{last}}.
\end{align}
Similarly, for a long-time evolution with $L^{QWC/FC}_{\text{max}}$ Trotter steps, noting that the last fragment in each Trotter step can be combined with the first one in the next Trotter step, results in an $R_z$ cost of,
\begin{align}
N_{R_z}^{QWC/FC}(\text{Long time}) &= 2N_H L^{QWC/FC}_{\text{max}} \notag \\ 
& - N^{QWC/FC}_{\text{last}}L^{QWC/FC}_{\text{max}} \notag \\ 
& - N^{QWC/FC}_{\text{first}}\left(L^{QWC/FC}_{\text{max}} - 1\right).  
\end{align}
Here $N^{QWC/FC}_{\text{first/last}}$ are the number of Pauli terms in the first/last fragment in the \gls{QWC}/\gls{FC} schemes respectively. These are expected to be significantly smaller compared to $N_H$, and thus the \gls{QWC} and \gls{FC} schemes will have the same asymptotic scaling for the cost of each Trotter oracle as the \gls{PF} scheme, which is also mentioned in the main text. However, the number of Trotter steps to perform the same total time evolution will be different between these three schemes, resulting in varied costs.

\section{Resource Estimates for Real Space Fragments} \label{app:rs_cost}
In this appendix we calculate the cost of implementing the Trotter fragments associated with the real space form. For simplicity our deduction here will not consider the coefficient caching technique, since otherwise the cost would depend on the specific structure of the non-zero Taylor coefficients in the Hamiltonian. We will consider an $n$-mode expanded Hamiltonian under a $d$-th degree Taylor expansion. Implementing a single $\ell$-th degree monomial, where $2\leq\ell\leq d$, will then require the following operations for implementing the associated circuit in Fig.~\ref{fig:rs_exponential}:
\begin{enumerate}
    \item Multiplication of the vibrational registers a total of $\ell-1$ times to produce a register with the $\ell$-th degree coefficient. Each vibrational register consists of $N_q$ qubits. Note that as multiplications happen, the new registers need to grow in size to represent the multiplication results. The cost of these multiplications will correspond to
\begin{equation}
        \mathcal C_1^{(\ell)}= \sum_{j=1}^{\ell-1} \mathcal{M}(N_q,jN_q),
\end{equation}
where we have defined $\mathcal{M}(a,b)=2ab-\max\{a,b\}$ as the T gate cost of multiplying two registers consisting of $a$ and $b$ qubits respectively. This procedure then requires $\mathcal{Q}_1^{(\ell)}= 2N_q+3N_q+\cdots+\ell N_q=N_q(\ell(\ell+1)/2-1)$ additional qubits and $\mathcal{C}_1^{(\ell)}=N_q^2\ell(\ell-1)-2N_q\ell+3N$ T gates.
    \item Loading of the Taylor coefficient in the register of $b_k$ qubits, which does not incur a T gate cost while still requiring the additional $\mathcal{Q}_2^{(\ell)}=b_k$ qubits.
    \item     Multiplication of the coefficient registers for a cost of $\mathcal{C}_3^{(\ell)} = \mathcal{M}(\ell N_q,b_k)=2\ell N_qb_k-\ell N_q$ T gates and additional $\mathcal{Q}_3^{(\ell)} = \ell N_q+b_k$ qubits.
    \item Addition operation for the phase gradient technique. Noting that the T gate cost of the addition operation is $\mathcal{A}(a)=4a-4$ when adding a register with $a$ qubits, the cost here will be given by $\mathcal{C}_4^{(\ell)}=4(\ell N_q+b_k)-4$.
    \item Uncomputation, consisting in re-applying the multiplications in steps 1-3, such that $\mathcal{C}^{(\ell)}_5 = \mathcal{C}^{(\ell)}_1+\mathcal{C}^{(\ell)}_3$.
\end{enumerate}
Overall this yields a T gate cost of 
\begin{align}
    \mathcal{C}^{(\ell)} &= 2\mathcal{C}^{(\ell)}_1 + 2\mathcal{C}^{(\ell)}_3+\mathcal{C}^{(\ell)}_4 \nonumber \\
    &=2N_q^2\ell(\ell-1)+6N_q+2N_q\ell(2b_k-1)+4b_k-4 \\
    &\sim\mathcal{O}(N_q^2 \ell^2)
\end{align}
per $\ell$-th degree coefficient. The maximum number of required additional qubits will be used when $\ell=d$, for which we have $\mathcal{Q}^{(d)}=N_q(d(d+1)/2-1)+b_k+d N_q+b_k=N_q (d^2+3d-2)/2+2b_k$.
Having derived the cost of implementing each coefficient in the Taylor expansion, we now proceed to counting the number of coefficients for the $n$-mode $d$-th degree Hamiltonian. For $\ell=1$ there are no terms to implement. For $\ell=2$, we can have up to $\binom{M}{2}$ terms associated to the $q_i q_j$ terms for $i\neq j$, and $\binom{M}{1}=M$ terms coming from the $q_i^2$ terms. For $\ell=3$, we then will have $\binom{M}{3}$ terms coming from $q_iq_jq_k$ terms for $i\neq j\neq k$, 2$\binom{M}{2}$ terms corresponding to $q_i^2 q_j$ and $q_iq_j^2$ kind of terms for $i\neq j$, and $M$ terms coming from $q_i^3$ terms. This trend will continue up until $\ell=n$, at which point the terms of the form $q_{i_1}q_{i_2}\cdots q_{i_\ell}$ get saturated. For e.g. $\ell=4$, the number of coefficients coming from two-variable monomials are associated to $q_i^3q_j$, $q_i^2q_j^2$, and $q_iq_j^3$. We thus have that in general, the number of $j$-variable monomials associated with a power $\ell$ can be obtained by multiplying $\binom{M}{j}$ by the number of combinations in which we can put $\ell-j$ indistinguishable particles in $j$ boxes, namely $\binom{\ell-1}{\ell-j}$. From this, we have that the number of possible terms with degree $\ell$ is
\begin{equation}
    \mathcal N^{(\ell)} = \begin{cases}
        0,\ \textrm{if}\  \ell < 2 \\
        \sum_{j=1}^\ell \binom{\ell-1}{\ell-j}\binom{M}{j},\ \textrm{if}\  2\leq\ell\leq n \\
        \sum_{j=1}^n \binom{n-1}{n-j} \binom{M}{j},\ \textrm{if}\ n<\ell\leq d.
    \end{cases}
\end{equation}
For simplicity we here assume that $d>n$, from which we can obtain the total cost of this approach as
\begin{align}
    \mathcal{C}_{\rm real\  space} &= \sum_{\ell=2}^{n} \mathcal{C}^{(\ell)}  \sum_{j=1}^\ell \binom{\ell-1}{\ell-j} \binom{M}{j} \nonumber \\
    &\ \ + \sum_{\ell=n+1}^{d} \mathcal{C}^{(\ell)} \sum_{j=1}^n \binom{n-1}{n-j} \binom{M}{j}.
\end{align}
Since the complexity is dominated by the $\binom{M}{n} \mathcal{C}^{(d)}$ term, we arrive to
\begin{equation}
\mathcal{C}_{\rm real\  space} \sim \mathcal{O}(d^2N_q^2 M^n).
\end{equation}
We have not included here the cost of implementing the kinetic energy operator with its associated \gls{QFT}s since the overall cost is dominated by the application of the $d$-th degree coefficients in the potential energy terms. Finally, we note that this complexity will be the same as for the implementation with the caching technique, having that the associated cost reduction will appear in the prefactors of this scaling.

\section{Decomposing Propagators for Bosonic Fragments into Pauli Rotations} \label{app:bf_mapping}

To implement the fragment propagators for the \gls{BF} scheme as a quantum circuit, we map the Bogoliubov unitaries $\mathcal{B}_{\nu}$ and diagonal fragment unitaries $e^{-iD_{\nu}t}$ in Eq.~\eqref{eq:prop_bf} onto qubit operators and then decompose them into products of Pauli rotations. The quadratic Gaussian unitary for each fragment can be decomposed via a Bloch-Messiah decomposition as \cite{braunstein2005,houde2024}
\begin{align}
    \mathtt{G}_{\nu}=  \ \mathtt{BS}^{(\nu)}  \mathcal{S}^{(\nu)} \ \mathtt{BS}^{(\nu)}.
    \label{eq:bog_uni}
\end{align}
The elementary Gaussian gates, beam-splitter ($\mathtt{BS}$) and one-mode squeezing ($\mathcal{S}$) are defined as \cite{liu2024}
\begin{align}
    \mathtt{BS}^{(\nu)} & =  e^{{\sum_{l>m}\delta_{lm}^{(\nu)} \left(b^{\dagger}_l b_m - b_l b^{\dagger}_m\right)}}, \label{eq:u_bs} \\
    \mathcal{S}^{(\nu)} & =e^{\sum_{l}\chi_{l}^{(\nu)}\left(b^{\dagger}_l b^{\dagger}_l -b_l b_l\right)} \label{eq:u_sq}.
\end{align}
The parameters in $\mathtt{BS}^{(\nu)}$ and $\mathcal{S}^{(\nu)}$ can be obtained from the parameters $\{\alpha_{lm}^{(\nu)},\beta_{lm}^{(\nu)}\}$ with details provided in Ref.~\cite{malpathak2025}. The two beam-splitters in Eq.~\eqref{eq:bog_uni} generally have distinct parameters.

The unitary corresponding to the quadratic diagonal fragment gets mapped as (ignoring the constant terms in Eq.~\eqref{eq:diag_Hlq})
\begin{align}
    e^{-iD_{0}t} &= e^{-i\sum_{l}\epsilon_{l} n_{l}t} = \prod_{l} e^{-i \epsilon_{l} n_{l}t} \notag \\
    & =\prod_{l} \prod_{\alpha_l} e^{-\frac{i}{2}  \epsilon_{\alpha_l} \alpha_l \left(\bm{1}-z_{\alpha_l}\right)t}. \label{eq:quad_unitory}
\end{align}
Similarly, the unitary corresponding to diagonal quartic fragments are mapped to qubits as 
\begin{align}
    e^{-iD_{\nu}t} &= e^{-i\sum_{lm} \eta_{lm}^{(\nu)} n_l n_m t} = \prod_{lm} e^{-i  \eta_{lm}^{(\nu)} n_l n_mt} \notag \\
    & =\prod_{lm} \prod_{\alpha_l\beta_m} e^{-\frac{i}{8} \eta_{lm}^{(\nu)}  \alpha_l\beta_m  {\left( \bm{1} - z_{\alpha_l} \right) \left( \bm{1} - z_{\beta_m} \right)}t}. \label{eq:quartic_unitory}
\end{align}
The corresponding mapping for the generator of beam-splitter unitary in Eq.~\eqref{eq:u_bs} is 
\begin{align*}
    b^{\dagger}_l b_m -& b_l b^{\dagger}_m = \sum_{\alpha_l\beta_m=0}^{N_l-2} \frac{\sqrt{\alpha_l+1}\sqrt{\beta_m+1}}{8} \nonumber \\ \Big[ 
    &\left( x_{\alpha_l} x_{\alpha_l+1} + y_{\alpha_l} y_{\alpha_l+1} \right) 
          \left( x_{\beta_m} x_{\beta_m+1} + y_{\beta_m} y_{\beta_m+1} \right) 
     - \nonumber \\ & \left( x_{\alpha_l} y_{\alpha_l+1} - y_{\alpha_l} x_{\alpha_l+1} \right) 
              \left( x_{\beta_m} y_{\beta_m+1} - y_{\beta_m} x_{\beta_m+1} \right) \Big].
\end{align*}
These terms commute with each other, which simplifies the exponentiation process. Therefore, the decomposition into elementary gates can be performed in a straightforward manner. The beam-splitter unitary is decomposed as follows,
\begin{align}
    \mathtt{BS}^{(\nu)} &=  e^{-it{\sum_{l>m}\delta_{lm}^{(\nu)} \left(b^{\dagger}_l b_m - b_l b^{\dagger}_m\right)}}\notag \\ &= \prod_{l>m} \prod_{\alpha_l\beta_m} \prod_{j=1}^8 e^{-i\zeta_{\alpha_l\beta_m}^{\nu,(lm)} P^j_{\alpha_l \beta_m} t }  
    \label{eq:beam_unitory}
\end{align}
where $P^j_{\alpha_l \beta_m}$ are Pauli words and $\zeta_{\alpha_l\beta_m}^{\nu,(lm)}$ are parameters obtained from $\delta_{lm}^k$. Similarly, the generator for the single-mode squeezing unitary in Eq.~\eqref{eq:u_sq} maps as
\begin{align*}
b^{\dagger}_l  b^{\dagger}_l &+ b_l  b_l =\\ &\sum_{\alpha_l=0}^{N_l -3}\frac{\sqrt{(\alpha_l+1)(\alpha_l+2)}}{2}  \left[ x_{\alpha_l} \ x_{\alpha_l+2} +  y_{\alpha_l} \ y_{\alpha_l+2} \right].
\end{align*}
However, these terms do not fully commute when $N_l > 3$. To address this issue, we introduce additional terms to ensure closure under commutation~\cite{izmaylov2020}. This modification simplifies both exponentiation and the elementary gate decomposition. The single-mode squeezing unitary
\begin{align}
    \mathcal{S}^{(\nu)}&=e^{-it\sum_{l}\beta_{l}^{(\nu)} \left(b^{\dagger}_l b^{\dagger}_l - b_l b_l\right)}
    \notag \\ &= \prod_{l} \prod_{\alpha_l} \prod_{j=1}^2 e^{-it \theta_{\alpha_l}^{(\nu),l} P^j_{\alpha_l}},
    \label{eq:squeezing_unitory}
\end{align}
where $P^j_{\alpha_l}$ are Pauli strings and $\theta_{\alpha_l}^{(\nu),l}$ are parameters obtained from $\chi_{l}^k$. The operator for the displacement unitary is mapped to qubits as
\begin{align*}
b^{\dagger}_l   +   b_l = \sum_{\alpha_l=0}^{N_l-2} \frac{\sqrt{\alpha_l+1}}{2} \left[ x_{\alpha_l} x_{\alpha_l+1} +  y_{\alpha_l} y_{\alpha_l+1} \right].
\end{align*} 
Similarly to the squeezing unitary, the terms in displacement unitaries also do not commute fully for $N_l > 2$, so we introduce additional terms to ensure closure under commutation. The displacement unitary can then be decomposed as follows:
\begin{align}
    \mathtt{D}^{(\nu)} &=e^{-it\sum_{l}\gamma_{l}^{(\nu)} \left( b_l - b^{\dagger}_l \right)}
    \\ &= \prod_{l} \prod_{\alpha_l} \prod_{j=1}^2 e^{-it \xi_{\alpha_l}^{{(\nu)},l} P^j_{\alpha_l}},
    \label{eq:displacement_unitory}
\end{align}
where $P^j_{\alpha_l}$ are Pauli strings and $\xi_{\alpha_l}^{{(\nu)},l}$ are parameters obtained from $\gamma_{l}^{(\nu)}$.

\section{Resource Estimates for Bosonic Fragmentation Scheme} \label{app:bf_res_est}
Here, we detail the procedure for calculating the $R_z$ count required to implement a single Trotter oracle, as well as for long-time evolution involving $L_{\mathrm{max}}$ Trotter steps within the \gls{BF} scheme. As described in Sec.~\ref{sec:cform_resource}, we apply the same procedure to estimate the number of T gates per $R_z$ gate. The Hamiltonian is partitioned into one quadratic fragment and $N_f$  quartic fragments, ordered such that the quadratic fragment appears first, followed sequentially by the quartic fragments. In this work, we assume that each mode is represented using the same number of modals, setting $N=N_l$  for all $l\in [1,M]$. When the Taylor expansion is truncated at a degree $d=4$, the estimated number of $R_z$ gates required per Trotter oracle is calculated from the following three contributions
\begin{enumerate}
    \item $MN$ for the diagonal quadratic terms.
    \item $\left[ \frac{M(M-1)}{2} N^2 + M\frac{ N (N-1)}{2} + M N \right] $ for the diagonal quartic terms.
    \item $ \left[ 4 M(M-1)(N-1)^2 + M(N-1) + M N \right] $ for the unitary operators.
\end{enumerate}
Combining these contributions and employing the second-order Trotter approximation, the total estimate for $R_z$ gate counts per Trotter oracle is given by
\begin{multline}
N_{R_z}^{\text{BF}}(\text{Trotter oracle}) = 2M N  \\ +   \left[ \frac{M(M-1)}{2} N^2 + M \frac{ N (N-1)}{2}  
+  M N \right] (2N_f-1) \\ + 2 \big[ 4  M(M-1) N^2
+ M (N - 1) +  M N \big] (N_f+1)  .
\end{multline}
Additional $R_z$ gates may arise from the inclusion of compensating terms introduced to enforce closure in the displacement and squeezing unitaries. However, these contributions are relatively minor compared to the dominant costs and are therefore neglected in this estimate. Note that we have applied simplifications by utilizing the group structure of the Bogoliubov unitaries. Specifically, we have taken two identities, first $\mathcal{B}_{N_f}^\dagger \mathcal{B}_{N_f} =\mathbb{I} $ and $\mathcal{B}_{\nu}^\dagger \mathcal{B}_{\nu+1} = \mathcal{B}^{\prime}$. The quartic diagonal contribution is the dominant cost, which scales as $\sim \mathcal{O}( M^2N^2 N_f)$.

For a Taylor expansion of degree $d=6$, the sextic bosonic Hamiltonian can be decomposed into fast-forwardable fragments, one of which takes the form,
\begin{align}
    H_{\nu} & = \mathcal{B}_{\nu} \left(\sum_{lmk}^{M} \chi^{(\nu)}_{lmk} {n}_l {n}_m {n}_k \right) {\mathcal{B}_{\nu}}^{\dagger} \label{eq:sol-bos-frags-d6}.
\end{align}
The structure of the Bogoliubov unitary $\mathcal{B}_{\nu}$ in this decomposition remains the same as described in Eq.~\eqref{eq:sol-bos-frags}. In this case, the dominant computational cost arises from the sextic diagonal term, which scales as $\sim \mathcal{O}( M^3N^3 N_f)$. Finally, for a general $d$th-order Taylor expansion, the asymptotic cost per Trotter oracle scales as $\sim \mathcal{O}( M^{\left\lceil\frac{d}{2}\right\rceil} N^{\left\lceil\frac{d}{2}\right\rceil}N_f)$.

For long-time evolution involving $L_{\mathrm{max}}$ Trotter steps, the $R_z$ cost is reduced by merging unitaries at the boundaries between consecutive Trotter oracles. It comprises three distinct contributions.
$(L_{\mathrm{max}}+1)$ number of diagonal quadratic terms, 
$2 (N_f-1)L_{\mathrm{max}}$ diagonal quartic terms, and
$2 (N_fL_{\mathrm{max}}+1)$ unitary operators. Summing these contributions yields the total $R_z$ gate count for the full evolution,
\begin{multline}
N_{R_z}^{\text{BF}}(\text{Long-time}) = MN(L_{\mathrm{max}}+1) + \\  2\left[ \frac{M(M-1)}{2} N^2 + M\frac{ N (N-1)}{2} + M N \right] (N_f-1)L_{\mathrm{max}} \\ + 2 \big[ 4 M(M-1)(N-1)^2 + M(N-1) \\+ M N \big] (N_fL_{\mathrm{max}}+1) .
\end{multline}

\end{document}